\documentclass[aps,prd,twocolumn,superscriptaddress,nofootinbib]{revtex4-2}

\usepackage{graphicx}
\usepackage{amsmath,amssymb,bm}
\usepackage{hyperref}
\usepackage{xcolor}
\usepackage{comment}
\usepackage{textcomp}
\usepackage{tikz}
\usepackage{pgfplots}
\pgfplotsset{compat=1.18}
\usetikzlibrary{positioning,arrows.meta,fit,calc,backgrounds}

\newcommand{\pr}{p_r}
\newcommand{\ph}{p_\varphi}
\newcommand{\Ord}{\mathcal{O}}
\newcommand{\Ltwo}{L^2}

\begin{document}

\title{Gravity-Informed Neural Networks for Post-Newtonian Binary Dynamics}

\author{G.~Barbagallo}
\email{gabriele.barbagallo@estudiante.uam.es}
\affiliation{%
\itshape\small
Instituto de F\'{\i}sica Te\'orica UAM/CSIC\\
C/ Nicol\'as Cabrera, 13--15, C.U.~Cantoblanco, E-28049 Madrid, Spain
}

\author{J.~Matulich}
\email{jmatulichfabres@gmail.com}
\affiliation{%
\itshape\small
Instituto de F\'{\i}sica Te\'orica UAM/CSIC\\
C/ Nicol\'as Cabrera, 13--15, C.U.~Cantoblanco, E-28049 Madrid, Spain
}

\date{\today}

\begin{abstract}
We introduce a gravity-informed neural-network (GravINNs) framework for learning post-Newtonian binary orbital dynamics. The method incorporates the equations of motion directly into the training phase and is developed at three complementary levels. First, we construct single-orbit surrogates for conservative post-Newtonian dynamics and assess their accuracy across different orbital configurations. 
We then extend the approach to a parametric model spanning a four-dimensional parameter space defined by the initial separation, radial and tangential momenta, and symmetric mass ratio, allowing a single trained network to represent entire families of orbits and replace repeated numerical integrations.
Finally, we consider dissipative dynamics by including the leading radiation-reaction contribution at $2.5$PN order. In this regime, the network reproduces the orbital evolution and the associated secular energy loss, starting from a conservative 2PN warm start and without using dissipative reference trajectories in the training.  These results show that physics-informed  neural-networks (PINNs) can provide accurate and flexible representations of both conservative and dissipative post-Newtonian dynamics. The complete implementation and the models used in this work are publicly available through the \texttt{GravINNs} repository.
\end{abstract}

\maketitle
\section{Introduction}
\label{sec:intro}

Since the first detection of gravitational waves from a merging binary
black hole~\cite{Abbott2016}, extracting astrophysical information from a
signal has required comparing it against theoretical waveforms across a
high-dimensional space of orbital parameters -- masses, spins,
eccentricity, orientation. Every such waveform rests on a solution of the
relativistic two-body problem, obtained from numerical
relativity~\cite{Pretorius2005}, effective-one-body
resummation~\cite{Buonanno:1998gg}, post-Newtonian (PN)
expansions~\cite{Blanchet2014}, post-Minkowskian
methods~\cite{Bern2019}, or gravitational self-force
theory~\cite{BarackPound2018}. Producing enough such solutions to cover
the parameter space densely is a central computational bottleneck, and has
driven an extensive effort on phenomenological waveform
families~\cite{Pratten2020} and on reduced-order and surrogate
models~\cite{Field2014,Blackman2017}.
 
Physics-informed neural networks (PINNs)~\cite{Raissi2019, Greydanus2019,Karniadakis2021}
were developed for exactly this kind of nonlinear problem: instead of
fitting precomputed solutions, a network is trained so that its output
satisfies the governing differential equation directly, with data entering
only as a sparse anchor. The method has been applied to fluid dynamics,
elasticity, and reaction-diffusion systems. 
Recent work has also begun to explore physics-informed and neural approaches in relativistic gravity, including PINN solutions of the Hamiltonian constraint and the vacuum Einstein equations, neural representations of spacetimes, physics-informed learning of PN corrections from simulation data, and structure-preserving neural flow maps for PN inspirals~\cite{Zhou:2026haz,Bayona2026, Cranganore:2025vsu,Yoo:2025yhl,GomesDaSilva2026}.

To our knowledge, this paper presents the first physics-informed neural surrogate that solves the post-Newtonian two-body dynamics directly across a continuous parameter space.
The comparison with existing tools should be
stated plainly, since the problem is already solved for a single trajectory: an adaptive Runge-Kutta (RK) integrator
returns one orbit faster, and to far higher accuracy, than any network we
could train. What a network offers instead is coverage of the
\emph{family} of orbits. Conditioned on the physical parameters, a single
trained network represents a continuous region of parameter space and
returns any member of it at the cost of a forward pass, without
retraining; and because the network is trained against the equations of
motion themselves, they act as a regularizer everywhere in that space, not
only near the sparse points where a reference orbit was computed.

We develop this approach on the conservative 2PN two-body problem, a
setting in which a high-accuracy numerical reference solution is available everywhere and every design
choice can be checked against it, and then extend it to the dissipative,
radiation-reaction sector. Writing the dynamics as an autonomous system in
the orbital angle, in place of time, lets us build much of the physics
directly into the network rather than into the loss: the initial data are
enforced exactly, angular momentum conservation is exact by construction,
and the radial periodicity of the orbit is represented by a bank of
harmonic features whose frequency is itself learned instead of being fixed by
hand. This formulation also exposes a limitation of physics-only training
that is not specific to gravity: because the equations are autonomous,
their residual is exactly invariant under a rigid rotation of the
solution, and cannot by itself select which member of that one-parameter family is the
physical one. Only the initial condition does, and since it is imposed at a
single angle its pull is weak when the orbit is nearly recurrent. The equations fix the shape of the orbit; sparse data or, more efficiently, a warm
start transferred from the analogous lower-order (1PN) solution make the correct
phase reliably found. This transfer does not in general make the network more
accurate; it changes, in a controlled region of the parameter space, whether the
correct solution is found at all.

Building on this formulation, we train a single network conditioned on
four physical parameters (initial separation, radial and tangential
momenta, and symmetric mass ratio) that reproduces the conservative 2PN
dynamics, reaching, with $K=2\times10^{4}$ anchor orbits, a
median relative orbit error of $3.7\times10^{-4}$ against Runge-Kutta
numerical integration, with $95\%$ of held-out orbits below $10^{-3}$.
At this anchor density a purely supervised network with the same architecture
already reaches $93.9\%$; the physics loss matters most at lower density
($K\le5\times10^{3}$, Sec.~\ref{sec:supervised}).
Analyzing these errors across the parameter space shows that they are not
uniform: the surrogate is accurate throughout the region where the PN
expansion is reliable, and its failures concentrate at the two extremes of the
orbit---close periapsis passages in the strong field, where the physics
residual is also most poorly conditioned, and the distant apoapsis of strongly
eccentric orbits. Since the reference is itself a 2PN solution, the breakdown of the expansion cannot by itself cause this disagreement; a comparison with a purely supervised model, which also degrades there, shows that it reflects the numerical difficulty of the region rather than the physics loss alone. The same architecture extends to the radiation-reaction inspiral at 2.5PN order without ever being shown a dissipative reference trajectory. In this regime, the pointwise residual is too small relative to the conservative dynamics to resolve the secular energy loss on its own, so we supplement it with an energy-balance residual that makes the radiation-reaction signal visible to the loss.
 
We regard the conservative two-body problem treated here as a deliberately
chosen first stage of a longer program, not as an attempt to replace established numerical methods for a problem they already solve well. The problem is rich enough to expose the main difficulties of physics-informed learning in gravitational dynamics, including  a residual blind to a physical
degree of freedom and a loss unable to resolve a small secular effect, while still remaining simple enough that these limitations can be identified clearly and the results can be checked against numerical reference solutions. These are, in our view, the necessary
lessons before extending physics-informed networks to settings where no
such reference exists, up to and including the field equations of general
relativity.

The paper is organized as follows. In Sec.~\ref{sec:formulation} we introduce the PN two-body dynamics in the angle domain and discuss the phase degeneracy of the autonomous formulation. Sec.~\ref{sec:single} presents the single-orbit surrogate, including its extension to long integrations. In Sec.~\ref{sec:architecture}
 we construct the parametric surrogate. Sec.~\ref{sec:dissipative}
 extends the framework to dissipative dynamics at 2.5PN order for a set of specific points in parameter space. Finally, Sec.~\ref{sec:discussion}
 summarizes the main findings and discusses possible outlooks.

A short account of the main results is given in a companion Letter~\cite{PRLcompanion}. The complete Python implementation \texttt{GravINNs} used in this work is publicly available~\cite{GravINNscode}. It is worth mentioning that all neural network models presented in this manuscript were trained locally using a mobile NVIDIA GeForce RTX 4090 GPU with 16 GB of VRAM.

\section{Post-Newtonian dynamics in the angle domain }
\label{sec:formulation}

\subsection{Reduced Hamiltonian }
\label{sec:Reduced Hamiltonian}

We work with the center-of-mass ADM Hamiltonian of the two-body problem in ADM coordinates, in the reduced form of Damour, Jaranowski and
Sch\"afer~\cite{Buonanno:1998gg,DJS2000,DJS2001}. We define the total mass $M=m_1+m_2$, the reduced mass $\mu=m_1 m_2/M$ and the symmetric mass ratio $\nu=\mu/M\in(0,1/4]$. The motion is confined to a plane with polar coordinates $(\tilde r,\varphi)$ and conjugate momenta $(\tilde p_r,\tilde p_\varphi)$; we write $\tilde p_\perp\equiv\tilde p_\varphi/\tilde r$ for the tangential component of the momentum. We introduce a reference separation $R_0$, the dimensionless radius $r\equiv\tilde r/R_0$, and the characteristic velocity $v_\ast\equiv\sqrt{GM/R_0}$, and define the dimensionless momenta $p_r\equiv\tilde p_r/(\mu v_\ast)$, $p_\varphi\equiv\tilde p_\varphi/(\mu R_0 v_\ast)$ and $p_\perp\equiv\tilde p_\perp/(\mu v_\ast)=p_\varphi/r$. The dimensionless speed of light is $c\equiv\tilde c/v_\ast$, with $\tilde c$ the speed of light in m/s. The Hamiltonian is normalized as $\hat H\equiv\tilde H/(\mu v_\ast^2)$ with the rest mass subtracted, and time is measured in units of $R_0/v_\ast$.
The reduced Hamiltonian through 2PN order expressed in normalized variables is then
\begin{equation}
\hat H \;=\; H_{\rm N} \;+\; \frac{1}{c^2}H_{\rm 1PN}
\;+\; \frac{1}{c^4}H_{\rm 2PN} ,
\label{eq:H}
\end{equation}
with $p^2 \equiv \mathbf{p}^2 = p_r^2 + p_\perp^2 = p_r^2 + {p_\varphi^2}/{r^2}$ and 
\begin{align}
H_{\rm N} &= \frac{p^2}{2}-\frac{1}{r},
\label{eq:HN}\\[2pt]
H_{\rm 1PN} &= \frac{1}{8}(3\nu-1)\,p^4
  -\frac{1}{2r}\Big[(3+\nu)p^2+\nu \pr^2\Big]
  +\frac{1}{2r^2},
\label{eq:H1}\\[2pt]
H_{\rm 2PN} &= \frac{1}{16}\big(1-5\nu+5\nu^2\big)p^6 \nonumber\\
 &\quad+\frac{1}{8r}\Big[(5-20\nu-3\nu^2)p^4-2\nu^2 p^2\pr^2-3\nu^2\pr^4\Big]
 \nonumber\\
 &\quad+\frac{1}{2r^2}\Big[(5+8\nu)p^2+3\nu \pr^2\Big]
 -\frac{1}{4r^3}\big(1+3\nu\big).
\label{eq:H2}
\end{align}

The truncation~\eqref{eq:H} carries its own domain of validity, which will
prove relevant to the results reported below. Defining $\tilde{r}_g=GM/\tilde{c}^2$, we adopt $\tilde{r}_{\rm bd}=20\tilde{r}_g$ as a conservative, heuristic label for the region where the 2PN truncation is expected to be reliable; no sharp threshold exists, since the accuracy depends on the mass ratio, the orbital state and the quantity considered (see, e.g., Ref.~\cite{Blanchet2014}). For each configuration we estimate the periapsis at Newtonian order from the initial data, $\tilde r_p=R_0\,r_p$, with $r_p$ given in Eq.~\eqref{eq:rp} below. This estimate holds for any initial radius and radial momentum; for $p_{r,0}=0$ it reduces to $\tilde r_p=\tilde r_0\,\gamma^2/(2-\gamma^2)$ for a sub-circular start ($\gamma<1$) and to $\tilde r_p=\tilde r_0$ for $1\le\gamma<\sqrt2$, where $\gamma$ is the initial tangential velocity in units of the circular velocity (Sec.~\ref{sec:family}). All benchmark statistics below use this estimate rather than the minimum separation of the 2PN reference orbit. The resulting periapsis is then compared with $\tilde{r}_{\rm bd}$: configurations with $\tilde{r}_p>\tilde{r}_{\rm bd}$ are considered to lie within the PN reliable regime. This criterion is used as a physical-validity label rather than as a selection filter: the parametric surrogate of Sec.~\ref{sec:architecture} is trained and tested over the whole parameter box, including orbits with $\tilde{r}_p<\tilde{r}_{\rm bd}$, where it reproduces the 2PN dynamics as a mathematical system.

As Section~\ref{sec:validity} will show, errors concentrate at close periapsis
passages below $\sim10\,\tilde r_g$ and at the distant apoapsis of highly
eccentric orbits, but the surrogate still succeeds on most orbits between
$10$ and $20\,\tilde r_g$, so this baseline criterion is conservative. Since
all reference orbits are 2PN solutions, the label concerns the physical
validity of the target model, not the accuracy with which the surrogate
reproduces it; the strong-field errors of Sec.~\ref{sec:validity} are
therefore numerical in origin.

\subsection{Hamilton's equations in the angle-domain}

The Hamiltonian~\eqref{eq:H} is independent of $\varphi$, so
$\ph\equiv L$ is exactly conserved, and independent of time, so
$\hat H\equiv E$ is exactly conserved. For an orbit with $L>0$, $\varphi(t)$ is strictly monotonic ($\dot\varphi=\partial\hat H/\partial L
= L/r^2 + \Ord(c^{-2}) > 0$), and can replace time as the evolution
variable. Although time would appear to be the natural evolution parameter, we choose to use $\varphi$ as it offers computational advantages for the problem considered here. The corresponding time-domain formulation is currently under study~\cite{BarbagalloandMatulich} and represents a natural extension of the present approach towards full GR.

Writing $u\equiv 1/r$, Hamilton's equations reduce to the
first-order autonomous system 
\begin{equation}
\frac{du}{d\varphi}
 = -u^2\,
   \frac{\partial\hat H/\partial \pr}{\partial\hat H/\partial L},
\qquad
\frac{d\pr}{d\varphi}
 = -\,
   \frac{\partial\hat H/\partial r}{\partial\hat H/\partial L},
\label{eq:angle-system}
\end{equation}
with $L$ entering as a constant fixed by the initial data. At Newtonian
order Eq.~\eqref{eq:angle-system} reduces to the classical Binet equation governing Keplerian conic sections; while the PN corrections lead to a non-closed orbit with a secular periapsis advance, producing the characteristic precessing-rosette structure. Three properties of this formulation are particularly relevant for what follows: (i) the domain $\varphi\in[0,\varphi_{\max}]$ is compact and can be chosen identically for all bound orbits, independently of their potentially different orbital periods; (ii) $u(\varphi)$ and $p_r(\varphi)$ are smooth, bounded, and nearly periodic functions, making them well suited to spectral representations; and (iii) the right-hand sides of Eq.~\eqref{eq:angle-system} contain no explicit dependence on $\varphi$, so that the resulting system is  \emph{autonomous}.
As we will see in Sec.~\ref{sec:phase}, the last property has important  consequences for the physics-informed training. 


\subsection{The four-parameter family }
\label{sec:family}

We parametrize the initial data at $\varphi=0$ by
\[
\theta\equiv(\nu,r_0,p_{r,0},\gamma),
\]
where $r_0=r(0)$, $p_{r,0}=p_r(0)$, and $\gamma\equiv v_0/v_c(r_0)$ is the initial tangential velocity in units of the Newtonian circular velocity at $r_0$, $v_c=\sqrt{GM/\tilde r_0}=v_\ast/\sqrt{r_0}$. Thus $p_\perp(0)=\gamma/\sqrt{r_0}$, $\gamma=1$ corresponds to a circular start at Newtonian order, and the conserved angular momentum is
\begin{equation}
L=p_\varphi=r_0\,p_\perp(0)=\gamma\sqrt{r_0}.
\label{eq:ICmap}
\end{equation}
The Newtonian energy~\eqref{eq:HN} and eccentricity associated with these initial data are
\begin{equation}
E_{N,0}=\frac12\left(p_{r,0}^2+\frac{\gamma^2}{r_0}\right)-\frac{1}{r_0},
\qquad
e_{N,0}=\sqrt{1+2E_{N,0}L^2},
\label{eq:E0e0}
\end{equation}
and the corresponding Newtonian periapsis is
\begin{equation}
r_p=\frac{L^2}{1+e_{N,0}}.
\label{eq:rp}
\end{equation}
These quantities are used as \emph{a priori} filters for bound motion
($E_{N,0}<0$) and, in Sec.~\ref{sec:architecture}, to bound the network output
range on an orbit-by-orbit basis.

Throughout this section we will use $m_1=m_2=1.4\,M_\odot$, $\nu =1/4$, $r_0=1$, $p_{r,0}=0$, $\gamma=0.8$ and
$R_0=200\,$km, unless explicitly specified.

\subsection{Phase blindness of the autonomous residual}
\label{sec:phase}

Before introducing any network we record the  property (iii) of the
system~\eqref{eq:angle-system}, because it determines the design of
everything that follows. The system has the form
$\bm y'(\varphi)=\bm F(\bm y(\varphi))$ with $\bm y=(u,\pr)$ and $\bm F$ explicitly
independent of $\varphi$. It is therefore invariant under translations of $\varphi$: if $\bm y(\varphi)$ is a solution, so is
$\bm y_\delta(\varphi)\equiv\bm y(\varphi+\delta)$ for every constant
$\delta$, since
$\bm y_\delta'(\varphi)=\bm y'(\varphi+\delta)
=\bm F(\bm y(\varphi+\delta))=\bm F(\bm y_\delta(\varphi))$.
Therefore, the dynamics do not know where $\varphi=0$ is; only the initial
data select a member of the one-parameter family of rigidly rotated orbits.

This has important consequences for residual-based training. A loss $\ell_{\rm PDE}$ built from
the equations of motion---however it is weighted, sampled, or
optimized---\emph{cannot distinguish a solution from any of its phase
translates}, because every translate has identically zero residual. 
In contrast, a trajectory error, such as the relative $L^2$ error,
compares the predicted solution with a specific reference trajectory and
is therefore sensitive to their relative phase. For two trajectories
$\bm y(\varphi)$ and $\bm y_\delta(\varphi)$, we define the relative
$L^2$ error as
\begin{equation}
\mathcal E_{L^2}(\delta)
\equiv
\frac{
\left[
\int_0^{\varphi_{\max}}
\left\|
\bm y_\delta(\varphi)-\bm y(\varphi)
\right\|^2\,d\varphi
\right]^{1/2}
}{
\left[
\int_0^{\varphi_{\max}}
\left\|
\bm y(\varphi)
\right\|^2\,d\varphi
\right]^{1/2}
}.
\label{eq:L2_error}
\end{equation}
Table~\ref{tab:phase} illustrates this effect using a numerical Runge--Kutta solution of the 2PN system: for instance,  a
rigid shift of $0.1\,$rad leaves the residual unchanged at the $10^{-8}$ order 
while the relative $L^2$ error reaches $2.7\times10^{-2}$. 
If the network prediction shifts along the orbit by changing its phase, $\ell_{\rm PDE}$ does not change significantly. Consequently, the gradient of the physics loss provides no signal that would drive the prediction back towards the correct phase. In other words, the residual functional, taken without initial data, does not
determine the phase. The initial-value problem itself is well posed; the
difficulty is that the exact initial condition acts only at $\varphi=0$ and is
released over the gate width $\tau$. A rigidly rotated configuration violates it
only by $\|\bm y(\delta)-\bm y(0)\|$, which is small for nearly recurrent
(near-circular) orbits, so the optimizer can reach it at almost no cost.
\begin{table}[t]
\begin{ruledtabular}
\begin{tabular}{lcc}
$\delta$ [rad] & $\ell_{\rm PDE}$ & $\mathcal E_{L^2}(\delta)
$ \\
\hline
$0$     & $6.7\times10^{-8}$ & $0$ \\
$0.01$ & $6.6\times10^{-8}$ & $1.4\times10^{-3}$ \\
$0.02$  & $6.9\times10^{-8}$ & $5.5\times10^{-3}$ \\
$0.05$  & $6.0\times10^{-8}$ & $1.4\times10^{-2}$ \\
$0.1$  & $6.6\times10^{-8}$ & $2.7\times10^{-2}$ \\
\end{tabular}
\end{ruledtabular}
\caption{Phase blindness of the angle-domain residual, measured on a Runge--Kutta
2PN orbit (three
revolutions with initial parameters $\theta=(0.24,1.27,0.18, 0.81)$).}
\label{tab:phase}
\end{table}

The phase must therefore be determined by additional information, which can
come from the initial condition, supervised data, or from the architecture
used to propagate the initial-condition information along the solution. The
initial condition fixes the phase at $\varphi=0$, while supervised data fix
the phase at the locations where they are provided. If this information is
not sufficiently propagated along the trajectory, the prediction can drift
in phase without being penalized by the physics residual. The failure modes
discussed in Secs.~\ref{sec:single} and~\ref{sec:single-longorbit} will be
understood in this way.

\section{The single-orbit surrogate}
\label{sec:single}

We first introduce the method in its most fundamental setting: employing a neural network to represent a \emph{single} bound orbit of the system~\eqref{eq:angle-system} at a fixed parameter point $\theta$. Sections~\ref{sec:single-longorbit} and~\ref{sec:architecture} then extend this setting in two directions, first to long-time integrations of individual orbits and then to the representation of the four-parameter family of bound orbits parametrized by $\theta=(\nu,r_0,p_{r,0},\gamma)$.

Beyond being the natural starting point, the single-orbit problem is a controlled laboratory. Since the reference orbit can be computed numerically to high accuracy, the error of the network can be evaluated without significant contribution from numerical one. This allows us to address two questions that are central to the subsequent developments: \emph{when} does the physics-informed residual alone determine the orbit, and \emph{when} does it benefit from supervised data to supply the phase that Sec.~\ref{sec:phase} shows the residual cannot see? 
The answer, developed in Sec.~\ref{sec:single-nearcircular}, is that data are inessential to achieve accurate predictions across almost the entire physically valid domain but become important for \emph{reliability} in the near-circular limit, where the degeneracy is most severe.

\subsection{Network and output composition}
\label{sec:single-arch}

We represent the solution of Eqs.~\eqref{eq:angle-system} with a neural network $\mathcal{N}$ that takes the orbital angle $\varphi$ as its only physical input. Before being passed to the network, $\varphi$ is mapped to a fixed Fourier-feature embedding, $\mathbb{F}(\varphi)$, to provide the network with a representation better suited to the oscillatory character of the orbital dynamics, as we will describe in detail below. The network itself consists of 5 fully connected layers of width 128 and $\tanh$ activation functions. Its two outputs, $\big(\mathcal N_u(\varphi),\mathcal N_{\pr}(\varphi)\big)$,
are not used directly as predictions for $u$ and $\pr$, while are passed through a composition function,
$$\varphi \; \rightarrow\; \mathbb{F}(\varphi)  \;\stackrel{\mathcal N}{\longmapsto}\;
\big(\mathcal N_u(\varphi),\mathcal N_{\pr}(\varphi)\big)\longmapsto\; \big(u(\varphi), p_r(\varphi)\big),
$$
that fixes, by construction,
three essential properties that a free network output cannot guarantee: that $u$ and $\pr$
remain within the physical range of the orbit at every $\varphi$, that the
initial condition holds exactly, and that no optimization gradients are expended
enforcing either constraint.  The angular momentum $L$ is not predicted by the network: it is fixed by the initial data through Eq.~\eqref{eq:ICmap}, and its conservation is therefore imposed exactly. The composition that guarantees  all of these conditions are satisfied is
\begin{equation}
u(\varphi) = u_{\rm min}+(u_{\rm max}-u_{\rm min})\,
   \sigma\!\Big[a_0+g(\varphi)\,\mathcal N_u(\varphi)\Big],
\label{eq:u-comp}
\end{equation}
\begin{equation}
\pr(\varphi) = \pr^{\max}\,
   \tanh\!\Big[b_0+g(\varphi)\,\mathcal N_{\pr}(\varphi)\Big],
\label{eq:pr-comp}
\end{equation}
with $\sigma$ the logistic function and
\begin{equation}
g(\varphi)\equiv 1-e^{-\varphi/\tau},\qquad \tau\equiv 0.02\,\varphi_{\max}.
\label{eq:gate}
\end{equation} 
In the following, we outline the structure and motivations of this transformation.

\paragraph{Squashing onto a physical range.}
Early in training, when loss gradients are large, unconstrained network outputs can easily wander into unphysical regimes (such as negative values for $u$ or extreme values for the momentum $\pr$). To prevent this, we transform the raw network outputs and rescale them to the physically allowed ranges of $u$ and $\pr$. For $u$, we use a logistic sigmoid to restrict the output to the interval $[u_{\min},u_{\max}]$, while for the signed radial momentum $\pr$ we use a $\tanh$ function to restrict it to $[-\pr^{\max},\pr^{\max}]$. The determination of these intervals—derived from measured orbital excursions—is detailed in Sec.~\ref{sec:single-longorbit}. 

\paragraph{Enforcing the initial condition exactly.}
At $\varphi=0$ the composition must reduce to the known initial data $u_0$ and
$ p_{r,0}$, for \emph{every} value of the network weights, since this is the
only local information available to break the phase degeneracy of
Sec.~\ref{sec:phase}. This is achieved by choosing the offsets $a_0$
and $b_0$ inside the nonlinear functions so that $u(0)=u_0$ and $p_r(0)=p_{r,0}$: inverting Eq.~\eqref{eq:u-comp} and ~\eqref{eq:pr-comp} at $\varphi=0$ gives
\begin{align}
a_0 &=\log\frac{u_0-u_{\rm min}}{\,u_{\rm max}-u_0\,},
\label{eq:a0}  \\
b_0& =\operatorname{artanh}\!\left(\frac{ p_{r,0}}{\pr^{\max}}\right).
\label{eq:b0}
\end{align}
Both are computed once, from the known initial data, and they are
not trained. In this way, the initial
condition never appears as a term competing with the physics residual in the loss.

\paragraph{Releasing the initial condition away from $\varphi=0$.}
The initial condition is imposed exactly at $\varphi=0$, but the network must be free to modify the output for $\varphi>0$. We therefore introduce the gate $g(\varphi)$ in Eq.~\eqref{eq:gate}, which multiplies the network output before it enters the nonlinear transformation. Since $g(0)=0$, the network contribution vanishes at the initial point, independently of its value, and the offsets $a_0$ and $b_0$ are sufficient to impose the initial condition. As $\varphi$ increases, $g(\varphi)$ approaches unity and the network contribution gradually becomes dominant. With $\tau=0.02\,\varphi_{\max}$, this transition takes place over a small fraction of the angular domain. The same value of $\tau$ is used for all orbits and is chosen only to release the initial-condition constraint over a short interval while keeping the output smooth.
Taken together, Eqs.~\eqref{eq:u-comp}--\eqref{eq:gate} incorporate the known initial conditions and physical bounds directly into the network output. The network therefore only needs to represent the evolution of the unknown trajectory between $\varphi=0$ and $\varphi_{\max}$.

\paragraph{Fourier input embedding.}
While the preceding
operations constrain the network’s output, we must also
modify its input representation. Specifically, we map $\varphi$ into a fixed random Fourier-feature embedding before entering the network. We emphasize that this a fixed architectural coordinate transformation designed to address a well-known optimization pathology. A network fed the bare angle $\varphi$ learns slowly-varying structure readily but struggles with rapid oscillation, a well-documented pathology of networks with smooth activations~\cite{Tancik2020}. The fix is to hand the network oscillatory building blocks directly: rather than $\varphi$ itself, the first layer receives
\begin{equation}
\mathbb F(\varphi)=
\Big[
\hat\varphi,
\sin(b_j^{\rm L}\hat\varphi),
\cos(b_j^{\rm L}\hat\varphi),
\sin(b_k^{\rm H}\hat\varphi),
\cos(b_k^{\rm H}\hat\varphi)
\Big],
\label{eq:fourier-features}
\end{equation}
where $\hat\varphi\equiv \varphi/\varphi_{\max}$ is the normalized angle  and
the frequencies $b_j^{\rm L}$ and $b_k^{\rm H}$ are sampled from
Gaussian distributions according to
\begin{equation}
b_j^{\rm L}\sim\mathcal{G}(0,\sigma^2_{\rm L}),
\qquad
b_k^{\rm H}\sim\mathcal{G}(0,\sigma_{\rm H}^2),
\qquad
\sigma_{\rm H}>\sigma_{\rm L},
\end{equation}
where $\mathcal{G}(\mu,\sigma^2)$ denotes a Gaussian distribution with mean
$\mu$ and variance $\sigma^2$. The index ranges are $j=1,\dots,N_{\rm L}$ and $k=1,\dots,N_{\rm H}$, where we use
$N_{\rm L}=10$, $N_{\rm H}=16$, $\sigma_{\rm L}=3$ and $\sigma_{\rm H}=8$
throughout. The embedding~\eqref{eq:fourier-features} therefore has dimension
$1+2N_{\rm L}+2N_{\rm H}=53$, which sets the width of the input layer.  The two frequency bands provide oscillatory
features over different frequency ranges: the lower-frequency band helps
represent the broader variation of the orbital variables, while the
higher-frequency band allows for faster variations. 
These frequencies are randomly sampled once at initialization and kept fixed during training, thus they encode no information about
the particular orbit. The purpose of this transformation is to provide the network with oscillatory functions of different frequencies, which makes it easier to represent the oscillatory dependence of $u$ and $p_r$ on $\varphi$ rather than using the angle alone.

This construction is effective for integrations covering a few radial periods, but becomes less suitable as the integration interval increases. If the domain contains
$
N={\varphi_{\max}}/{\Phi_r}
$
radial periods, the radial motion undergoes $N$ repetitions over the normalized input interval. The random Fourier features do not contain the radial frequency $\Phi_r^{-1}$ as a prescribed scale, so the network must infer the corresponding recurrence from the available features. A small error in this frequency produces a phase error that grows with the number of radial periods. Increasing $\sigma_{\rm H}$ does not solve this problem, since adding higher-frequency random features does not ensure that the orbital frequency or its harmonics are represented accurately. To resolve this problem, Sec.~\ref{sec:single-longorbit} introduces a third feature bank built from harmonics of the orbit's \emph{own} radial frequency (learned during training), demonstrating that supplying this physical recurrence scale is what makes long integrations tractable.

\subsection{The physics loss}
\label{sec:single-loss}
The equations of motion~\eqref{eq:angle-system} can be written as two residual equations by bringing all terms to the left-hand side,
\begin{align}
R_u(\varphi) &= \frac{du}{d\varphi}
 +u^2\frac{\partial\hat H/\partial p_r}{\partial\hat H/\partial L},
  \label{eq:Ru}\\[2pt]
  R_{p_r}(\varphi) &= \frac{dp_r}{d\varphi}
+\frac{\partial\hat H/\partial r}{\partial\hat H/\partial L}.
  \label{eq:Rpr}
  \end{align}
  An exact solution of the equations of motion satisfies $R_u=R_{p_r}=0$ at every value of $\varphi$. In the RHS of Eqs.~\eqref{eq:Ru}--\eqref{eq:Rpr}, the
derivatives $du/d\varphi$ and $d\pr/d\varphi$ are computed by automatic
differentiation \emph{of the network's own output} through
Eqs.~\eqref{eq:u-comp}--\eqref{eq:pr-comp}. The remaining terms of the RHS are evaluated analytically  from the Hamiltonian at the phase-space point $(u,\pr,L)$ predicted by the network. The residual therefore measures whether the trajectory generated by the network satisfies the 2PN equations of motion, without requiring a reference numerical solution. This is what makes the training physics-informed rather
than supervised.

We enforce these equations at collocation points sampled uniformly across the angular domain, $\varphi_i\sim\mathcal U[0,\varphi_{\max}]$. At each training step, $N_c=4000$ points are newly sampled, allowing the equations to be enforced at different locations along the orbit. The physics loss is defined as
\begin{equation}
\ell_{\rm PDE}
=\frac{1}{N_c}\sum_{i=1}^{N_c}
\left[
R_u^2(\varphi_i)+R_{p_r}^2(\varphi_i)
\right],
\label{eq:pdeloss}
\end{equation}
and minimizing it drives the network toward a function that satisfies the equations of motion throughout the interval $[0,\varphi_{\max}]$.

Energy conservation provides an additional constraint that can be used during training. The energy residual is defined as 
\begin{equation}
R_E(\varphi) = \hat H(u(\varphi) ,p_{r}(\varphi) ,L)-E,
\end{equation}
where $\hat H(u(\varphi) ,p_{r}(\varphi) ,L)$ is evaluated analytically  at the phase-space point $(u, p_r , L)$ predicted by
the network and $E$, being conserved,  is computed exactly using the Hamiltonian evaluated at the initial condition. We quantify deviations from this energy shell with
\begin{equation}
\ell_E
=
\frac{1}{N_c}
\sum_{i=1}^{N_c}
R_E^2(\varphi_i).
\label{eq:eloss}
\end{equation}
In principle, this constraint is redundant because an exact solution of Hamilton's equations with the correct initial condition conserves the Hamiltonian automatically. In practice, however, the network does not satisfy the equations exactly during the early stages of training. Since $\ell_E$ is cheap to evaluate and involves no derivatives, we give it a weight larger than that of the physics residual, which keeps the prediction on the correct energy shell while the residual is still large. The energy term is therefore used as a numerical aid to the optimization, not as an independent physical constraint.

\subsection{Training Regimes: Pure Physics-Informed vs. Transfer Learning }
\label{sec:single-train}
We consider two training strategies for the single-orbit problem. Both use the physics loss introduced above, while they differ in whether supervised information from a reference orbit is used during training.
As an optional data-driven term, we introduce the loss
\begin{equation}
\begin{aligned}
\ell_{\rm anchor}
&=
\frac{1}{N_c}\sum_{i=1}^{N_c}
\Big[
\big(u(\varphi_i)-u^{\rm ref}(\varphi_i)\big)^2 \\
&\qquad\qquad+
\big(p_r(\varphi_i)-p_r^{\rm ref}(\varphi_i)\big)^2
\Big],
\end{aligned}
\label{eq:anchorloss}
\end{equation}
which measures the deviation of the network prediction from a single reference trajectory. The total loss is therefore
\begin{equation}
\ell
=
w_a\,\alpha\,\ell_{\rm anchor}
+
\beta\big(\ell_{\rm PDE}+\lambda_E\,\ell_{E}\big),
\label{eq:single-total}
\end{equation}
where $\alpha,\beta\in[0,1]$ are scheduled during training (see below), and the fixed weights $w_a=10$ and $\lambda_E=5$ set the relative scale of the anchor and energy terms with respect to the physics residual. For the long orbits of Sec.~\ref{sec:single-longorbit} we also add a term $\tfrac12\ell_E$ that is independent of $\beta$, so that the energy constraint remains active during the supervised warm-up stage, when $\beta=0$.

We use two limiting training modes to study the role of the anchor:
\begin{itemize}
\item \textbf{Physics-only (``cold'').} We set $\alpha=0$ and train the network from a random initialization using only the physics and energy losses. The only information about the desired trajectory is therefore the initial condition, which is enforced exactly by the network architecture.
\item \textbf{Transfer (``warm'').} Training is divided into three stages. In stage~1, the network is fitted  to a reference orbit that we choose to be the 1PN solution for the same initial data, with $\beta=0$. In stage~2, the physics loss is introduced gradually: $\beta$ is increased to one while $\alpha$ is decreased linearly to zero. In stage~3, the network is trained using the physics loss alone, with $\alpha=0$ and $\beta=1$.
\end{itemize}
The reference trajectory used for transfer is deliberately computed at 1PN order, using a Runge--Kutta integration, and therefore it is  not a solution of the 2PN equations enforced during training.
The purpose of the 1PN reference is not to provide the final solution, but rather to serve as a useful initial guess for the optimizer. Since it closely follows the 2PN orbit in both orbital structure and phase, it helps place the optimizer near the desired solution. The reference is then gradually removed during training rather than instantaneously, since, as we will see in Sec.~\ref{sec:single-longorbit}, an abrupt transition to the physics loss can lead to a loss of accuracy. The transfer procedure should therefore be viewed as an initialization strategy: the lower-order solution provides a starting point, while the final solution is determined entirely by the target 2PN dynamics. Optimization is performed using the Adam optimizer with a cosine learning rate schedule decaying from $3\times10^{-3}$ to $3\times10^{-6}$. We find that cosine annealing is important for reaching low final losses: with a constant learning rate, optimization can get trapped in sub-optimal local minima, with final losses several orders of magnitude larger than those obtained with the cosine schedule.

\subsection{Reliability and the near-circular degeneracy}
\label{sec:single-nearcircular}
We now focus on the reliability of physics-only training rather than only on its final accuracy. The two turn out to be different, and
the difference is a direct evidence we can offer for the relevance of
this study to numerical relativity. Across the PN-valid regime, a single orbit can generally be reproduced with an $\Ltwo$ error of order $10^{-5}$--$10^{-4}$ using the physics residual alone. The warm start does not in general improve the accuracy, except for the most circular orbits. However, accuracy alone does not reveal whether the training consistently finds the correct solution.
As discussed in Sec.~\ref{sec:phase}, the physics residual is invariant under shifts of the orbital phase. This becomes particularly relevant for nearly circular orbits, for which a small phase shift produces only a small change in the trajectory. The loss therefore develops an almost degenerate direction, making it possible for physics-only training to converge to a phase-shifted solution. The effect is small over a few radial periods but becomes more visible over longer integrations, so we use 10 radial periods throughout  this subsection. 
\paragraph{Protocol.}
We consider orbits with different initial eccentricities, spanning the near-circular to moderately eccentric case, keeping the periapsis  well inside the 2PN-valid regime ($\tilde{r}_p>\tilde{r}_{bd}$), so that difficulty is controlled only 
by degeneracy. For each orbit, both physics-only and warm-started training are performed with two independent random initializations. We define a failure as a converged run with $\mathcal{E}_{\Ltwo}>10^{-3}$.

\begin{figure*}[!]
    \centering
    \includegraphics[width=0.65\textwidth]{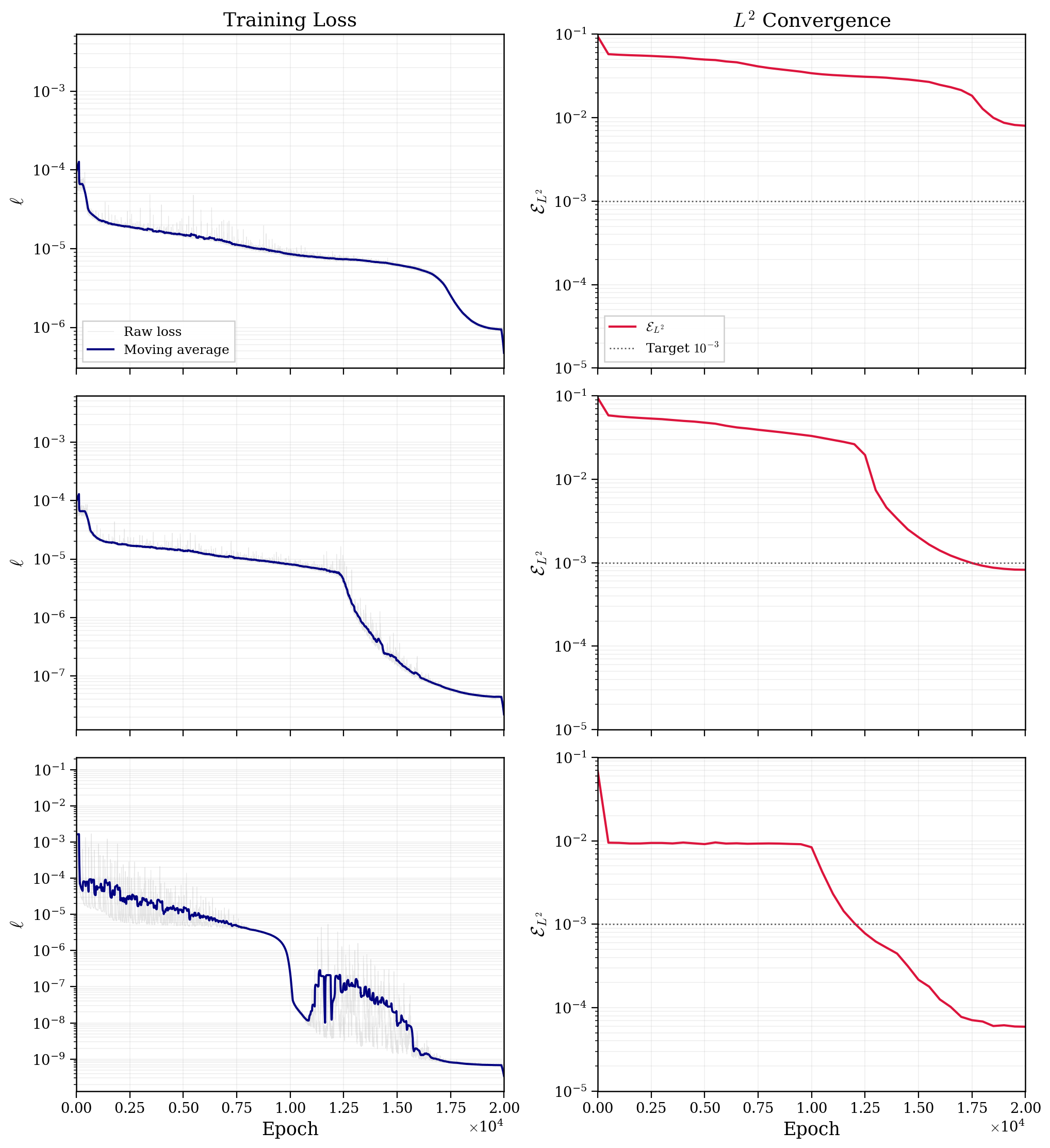}
 \caption{
Comparison of single-orbit training outcomes for different initialization strategies.
The left column shows the training loss, while the right column reports the corresponding convergence of  $\mathcal{E}_{L^2}$.
From top to bottom, the rows correspond to a failed cold start, a successful cold start, and a successful transfer-learning run.
The final errors are $\mathcal{E}_{L^2}=8.01\times10^{-3}$, $8.23\times10^{-4}$, and $5.92\times10^{-5}$, respectively.
The dotted line marks the target accuracy $\mathcal{E}_{L^2}=10^{-3}$.
}
    \label{fig:singleorbit}
\end{figure*}

\paragraph{Results.} 
The results are summarized in Table~\ref{tab:nearcircular}. Evaluated over 25 orbits in each eccentricity stratum, with two random initializations per orbit (50 runs per stratum), physics-only training fails in $38\%$ of the runs ($19/50$) for the most circular orbits ($e_0\simeq0.02$), while no failures are observed with the warm start. The failure rate falls with eccentricity: $10\%$, $6\%$, and $2\%$ at $e_0\simeq0.11$, $0.26$, and $0.54$, respectively. The loss of reliability is therefore strongly eccentricity-dependent, as expected from the phase degeneracy, but it is not strictly confined to the near-circular regime. Warm-started training did not fail in any of the $200$ runs performed. A representative comparison is shown in Fig.~\ref{fig:singleorbit}, where a failed cold start, a successful cold start, and a successful transfer-learning run are compared through their training loss and $\mathcal{E}_{L^2}$ convergence.

For the runs that do converge, the two approaches generally achieve comparable
accuracy. In the most circular case, however, the warm start improves both
reliability and accuracy, with median errors of $3.2\times10^{-5}$ compared
with $1.3\times10^{-4}$ for physics-only training. The fact that different
random seeds fail for different orbits indicates that the effect is associated
with the loss landscape rather than with a particular initialization.

\begin{table}[t]
\begin{ruledtabular}
\begin{tabular}{lcccccc}
 & \multicolumn{3}{c}{Physics-only (cold)} & \multicolumn{3}{c}{Transfer (warm)} \\
\cline{2-4}\cline{5-7}
$e_0$ & Fail & Med$\,\ell_{\rm PDE}$ & Med$\,\mathcal{E}_{\Ltwo}$
      & Fail & Med$\,\ell_{\rm PDE}$ & Med$\,\mathcal{E}_{\Ltwo}$ \\
\hline
$0.02$ & $38\%$ & $2.9\times10^{-8}$ & $1.3\times10^{-4}$
       & $0\%$   & $1.6\times10^{-9}$   & $3.2\times10^{-5}$ \\
$0.11$ & $10\%$   & $2.9\times10^{-8}$ & $5.7\times10^{-5}$
       & $0\%$    & $5.7\times10^{-9}$ & $8.8\times10^{-5}$ \\
$0.26$ & $6\%$    & $1.2\times10^{-7}$ & $8.8\times10^{-5}$
       & $0\%$    & $2.3\times10^{-8}$ & $3.7\times10^{-5}$ \\
$0.54$ & $2\%$    & $3.1\times10^{-7}$ & $1.5\times10^{-4}$
       & $0\%$    & $8.0\times10^{-8}$ & $1.4\times10^{-4}$ \\
\end{tabular}
\end{ruledtabular}
\caption{Reliability of physics-only versus warm-started training over ten
radial periods at $2$PN. ``Fail'' is the fraction of runs
reaching $\mathcal{E}_{\Ltwo}>10^{-3}$; each entry is based on 50 runs (25 orbits, two seeds each). Med$\ell_{\rm PDE}$ and Med$\mathcal{E}_{\Ltwo}$ are medians over the converged
runs. 
The
failing runs reach residuals within a factor of a few of the converged ones in
the same stratum, so failure does not indicate a lack of convergence.}
\label{tab:nearcircular}
\end{table}

The residual reached by the failing runs makes this fact explicit. Failure does not mean that training failed to
converge: these runs converge, and they converge to configurations that
satisfy the equations of motion while describing the wrong orbit. This is the
phase degeneracy of Sec.~\ref{sec:phase}.

\paragraph{Relevance to the fully relativistic problem.}
This example is also relevant to more general physics-informed models. In
numerical relativity, for example, minimizing only the evolution-equation
residual does not necessarily guarantee that the Hamiltonian and momentum
constraints are numerically satisfied~\cite{Gundlach2005}. Similarly, gauge
conditions introduce additional freedom that is not fixed by the evolution
equations alone. Although these cases are not identical to the phase degeneracy
considered here, they illustrate the same practical issue: a small physics
residual does not always guarantee that the optimizer has found the desired
physical solution.
In the present two-body problem the degeneracy is simple and appears mainly
near circular orbits. More complex relativistic systems can contain many
additional directions that are weakly constrained by the loss. This suggests
that the transfer strategy may become increasingly important as the
dimensionality of the problem grows. Work in this direction is currently under
way~\cite{BarbagalloandMatulich}.

\subsection{Long orbits: phase accumulation and the radial-harmonic
representation}
\label{sec:single-longorbit}
In this section we ask whether the architecture introduced in
Sec.~\ref{sec:single-arch} can represent the solution accurately over long integration intervals. In this case the main problem arises from the accumulation of small phase errors in the
network representation. 

The limitation becomes apparent when the architecture of
Secs.~\ref{sec:single-arch}--\ref{sec:single-train} is applied to longer
orbits. With this setup, the network reaches
$\mathcal{E}_{\Ltwo}\sim10^{-5}$ over 10 radial periods, but its performance gets worse
substantially when the integration interval is increased. For integrations
covering about $50$ revolutions, using 5 initial conditions with
$e_0$ in the range $[0.06,0.64]$ and safe periapses ($\tilde{r}_p>\tilde{r}_{bd}$), the  network
fails to reach the target accuracy, namely $\mathcal{E}_{\Ltwo}\sim10^{-3}$, in 4 of the 5 cases and stalls at
$\mathcal{E}_{\Ltwo}\sim10^{-2}$, while the loss reaches a plateau.
Increasing the training time does not remove this plateau, indicating that
the limitation is not simply due to insufficient optimization.

The origin of this degradation can be understood from the periodic structure
of the radial motion. For a bound conservative orbit, the radial variables
repeat after an angular interval $\Phi_r$ corresponding to two consecutive
periapsis passages. A network that represents the orbit over many revolutions
must therefore reproduce the same radial pattern repeatedly and with a very
small phase error. If the frequency represented by the network differs from
the true radial frequency by a fractional amount $\epsilon$, the resulting
phase error grows approximately as
$\Delta\varphi \sim \epsilon\varphi_{\max}$.
An error that is negligible over a few revolutions can therefore become
significant over tens of revolutions. The problem is particularly relevant
for the random Fourier features used in the architecture, since
their frequencies are sampled independently of the orbital frequency. Therefore, the
network must  construct the true periodicity from features
that do not explicitly contain the physical radial frequency. For long
integrations, this becomes a limitation of the representation itself.

\paragraph{Four modifications.}
We introduce four changes to the architecture and training procedure to
extend the surrogate to long orbits.

\emph{(i) Radial-harmonic features.}
The radial motion of a conservative bound orbit is periodic in the angular
domain,
$
u(\varphi+\Phi_r)=u(\varphi),
$
where $\Phi_r$ is the angular separation between successive periapsides.
The relativistic periapsis advance changes $\Phi_r$ from its Newtonian value
$2\pi$, but does not introduce a secular drift in the radial variables.
This known periodicity can therefore be incorporated directly into the
input representation. We supplement the random Fourier features of
Eq.~\eqref{eq:fourier-features} with
\begin{equation}
\mathbb{F}(\varphi)=
\left\{
\cos(n\omega\varphi),\,\sin(n\omega\varphi)
\right\}_{n=1}^{N_h},
\qquad
\omega=\frac{2\pi}{\Phi_r},
\label{eq:harmonic-bank}
\end{equation}
where $\omega$ is a  \emph{learnable} scalar\footnote{For a single orbit this is just a  number: $\omega$ is stored directly as a
trainable parameter of the network, with no auxiliary structure, since $\Phi_r$
can be measured from that orbit's own reference. Section~\ref{sec:architecture}
replaces it by a small auxiliary network once $\omega$ must become a function
of the orbital parameters.} initialized from the measured
periapsis spacing of the reference orbit.  The same set of harmonics is then
used at every radial cycle, so the representation does not have to learn a
new oscillatory pattern at each revolution. We use $N_h=10$, which is
sufficient for the orbits considered here.
This construction does not remove phase accumulation completely, since a
small error in the learned $\omega$ still produces a phase error that grows
with $\varphi_{\max}$. It does, however, make the relevant frequency an
explicit parameter of the representation rather than something that must be
reconstructed from a set of unrelated random frequencies. 

\emph{(ii) Output ranges from the measured orbital excursion.}
The output ranges in Eqs.~\eqref{eq:u-comp}--\eqref{eq:pr-comp} were initially
chosen using padded Newtonian estimates. These ranges can be much larger than
the actual deviation of a nearly circular orbit. As a result, the physical
trajectory occupies only a small fraction of the available output interval,
particularly for $p_r$. This degrades the conditioning of the output
transformation. For example, from
$p_r=p_r^{\max}\tanh(z)$, a perturbation in the network output gives
approximately 
$
\delta p_r\simeq p_r^{\max}\delta z
$,
when $z$ is small. Thus a large value of $ p_r^{\max}$   maps small errors in the raw network output, $\delta z$, into larger errors in the physical momentum.
We instead determine the output ranges from the measured variation of the
reference orbit and add a fixed relative padding. This keeps the trajectory
within a comparable fraction of the available range, with an occupancy of
roughly $70$--$80\%$ for the cases considered. The effect is particularly
large near circularity. For the most circular orbit tested, the original ranges left the residual
unable to fall below $2\times10^{-4}$---four orders of magnitude above its
value in well-conditioned runs---with a trajectory error of
$6.8\times10^{-2}$. With the ranges rescaled, the same configuration reaches
$\mathcal{E}_{\Ltwo}=3.6\times10^{-5}$.

\emph{(iii) Causal weighting with gradual removal.}
For long integrations, we introduce a causal weighting of the physics residual,
$
w(\varphi)=\exp[-\epsilon\,\Sigma(\varphi)],
$
where $\Sigma(\varphi)$ is the normalized cumulative residual up to $\varphi$: \begin{equation}
\Sigma(\varphi)
\equiv 
\frac{
\displaystyle \int_0^\varphi
\left[
R_u^2(\varphi')
+
R_{p_r}^2(\varphi')
\right]\,d\varphi'
}{
\displaystyle \int_0^{\varphi_{\max}}
\left[
R_u^2(\varphi')
+
R_{p_r}^2(\varphi')
\right]\,d\varphi'
}.
\label{eq:cumulative-residual}
\end{equation}
In practice $\Sigma$ is evaluated as a cumulative sum over collocation points
sorted in $\varphi$, normalized by the total over the batch so that
$\Sigma\in[0,1]$ and $\epsilon$ is independent of $\varphi_{\max}$ and of the
overall residual scale. The weight is largest near the initial condition and decreases as the
cumulative residual grows. The network is therefore forced to first
satisfy the equations close to $\varphi=0$ and then progressively extend the
solution to larger values of $\varphi$. This helps propagate the phase fixed
by the initial condition along the orbit.
The weighting is particularly useful for near-circular orbits, for which the
physics residual provides only weak information about the orbital phase.
However, removing the weighting abruptly can cause the network to move to a
different phase trajectory. We therefore decrease $\epsilon$ gradually during
training, so that the weighted residual continuously approaches the standard
unweighted residual. The final stage is performed with $\epsilon=0$.
For eccentric orbits, the effect is less pronounced because the sharp
periapsis passages provide a stronger phase reference.

\emph{(iv) Two-stage optimization.} Finally, the training is divided into a search stage and a refinement stage, both using the Adam optimizer. The first stage uses the cosine-annealed learning-rate schedule described in Sec.~\ref{sec:single-train}. At the end of this schedule, the learning rate reaches its lower bound, so simply continuing the schedule does not provide efficient refinement. We therefore restart the Adam optimizer from the state reached at the end of the search stage, using a smaller fixed learning rate for the second refinement stage. The restart rate is chosen between two extremes: rates that are too small give little additional improvement, whereas rates that are too large disturb the solution found during the first stage.

\paragraph{Result.}
\begin{figure*}[!]
    \centering
    \includegraphics[width=0.9\textwidth]{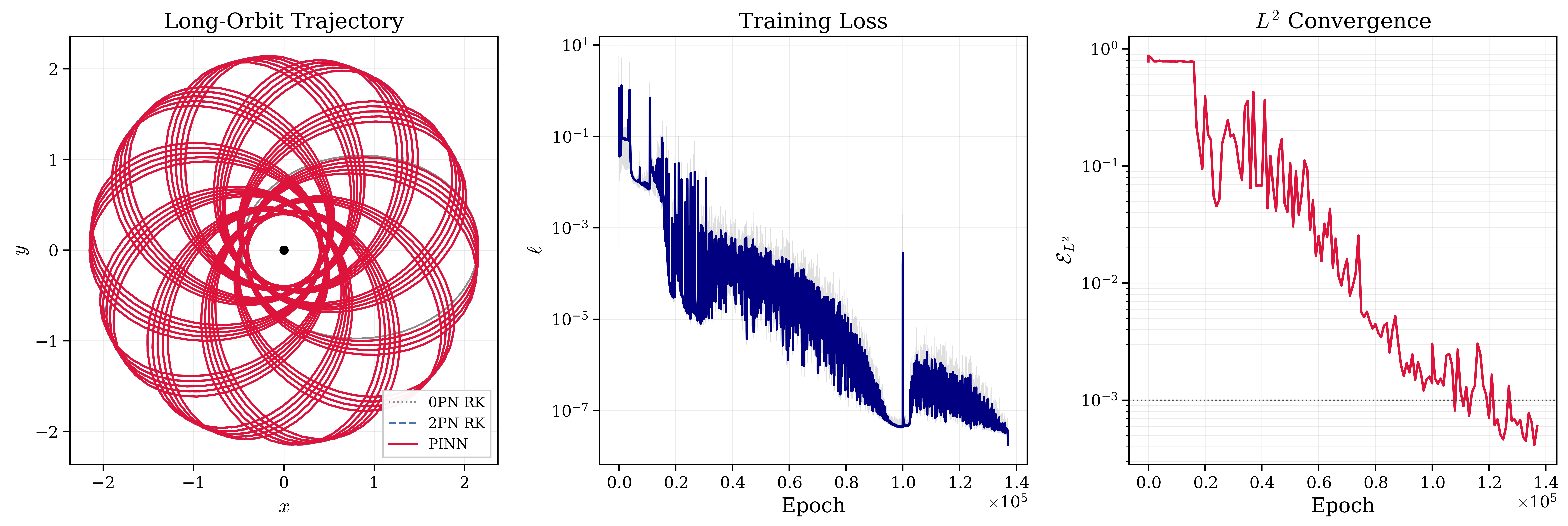}
    \caption{Long-orbit surrogate for a representative conservative 2PN configuration.
Left: orbital trajectory predicted by the PINN compared with the 0PN and 2PN Runge--Kutta solutions. 
Center: evolution of the training loss, with the raw history shown in light gray and a smoothed trend in blue. 
Right: convergence of the relative phase-space error $\mathcal{E}_{L^2}$ during training; the dotted line marks the target accuracy $\mathcal{E}_{L^2}=10^{-3}$.}
    \label{fig:longorbit}
\end{figure*}
With these four modifications, the surrogate reproduces all five test orbits
over approximately $50$ revolutions with $\mathcal{E}_{\Ltwo}<10^{-3}$ and a $2.3\times10^{-4}$ median error, as shown in
Table~\ref{tab:longorbit} and Fig.~\ref{fig:longorbit}. The first-stage search already reaches the target in all
five cases, while the refinement stage reduces the final error by factors of
1.1--4.5. Thus, the refinement improves the solution but is not
responsible for making the long integrations possible.
The final errors also show no clear dependence on eccentricity: they span a
factor of about 17 across the five orbits, with the most circular case
giving the smallest error. This is in contrast with the baseline
configuration, for which the error increased strongly towards the circular
limit. The comparison shows that part of the apparent difficulty of
near-circular orbits was caused by the output parametrization rather than by
the dynamics themselves. However, this result does not remove the phase degeneracy discussed in
Sec.~\ref{sec:single-nearcircular}. That experiment was repeated using the
corrected output ranges, and the failure rate in the most circular
eccentricity bin remained unchanged. The two effects should therefore be
distinguished: the output parametrization and radial-harmonic representation
determine how accurately a converged solution can be represented, whereas the
phase degeneracy determines whether optimization reaches the correct branch
in the first place.

\begin{table}[h]
\begin{ruledtabular}
\begin{tabular}{lcccc}
$e_0$ & $\tilde{r}_p/\tilde{r}_g$ & Revs & Search $\mathcal{E}_{\Ltwo}$ & Refine $\mathcal{E}_{\Ltwo}$ \\
\hline
$0.06$ & $56$ & $53.5$ & $3.9\times10^{-5}$ & $3.6\times10^{-5}$ \\
$0.19$ & $54$ & $52.5$ & $9.3\times10^{-4}$ & $6.3\times10^{-4}$ \\
$0.36$ & $24$ & $54.0$ & $3.1\times10^{-4}$ & $2.3\times10^{-4}$ \\
$0.51$ & $24$ & $53.2$ & $5.9\times10^{-4}$ & $4.8\times10^{-4}$ \\
$0.64$ & $23$ & $53.5$ & $9.3\times10^{-4}$ & $2.1\times10^{-4}$ \\
\end{tabular}
\end{ruledtabular}
\caption{Long-orbit test results over approximately 50 revolutions, showing the $\Ltwo$ error after the search and refinement stages.}
\label{tab:longorbit}
\end{table}

\paragraph{Beyond 50 revolutions.}
The four modifications extend the surrogate's range to roughly 50 revolutions, but accuracy degrades smoothly over longer intervals. For the $e_0=0.64$ case, the error increases from $2.1\times10^{-4}$ at 50 revolutions to $3.9\times10^{-2}$ at 100 revolutions.

Extensive testing confirms this is a structural limitation: drastically increasing collocation density, adding refinement cycles, or refining the $\Phi_r$ measurement fails to eliminate this error floor. Doubling the network capacity reduces the terminal residual by a factor of seven but improves $\Ltwo$ by only 12\%. 

This points directly to the phase degeneracy discussed in Sec.~\ref{sec:phase}. A larger network satisfies the autonomous equations more accurately, but this does not prevent small phase mismatches from accumulating over many radial cycles. Consequently, 50 revolutions represents the practical limit of the current architecture. Extending this range will likely require hard-coding radial periodicity directly into the harmonic representation, providing a stronger global constraint than the localized initial-condition gate of Eq.~\eqref{eq:gate}.


\section{The parametric surrogate}
\label{sec:architecture}
The single-orbit surrogate introduced in Sec.~\ref{sec:single} represents the
solution corresponding to one fixed point in the parameter space
$\theta=(\nu,r_0,p_{r,0},\gamma)$. We
now extend this approach, training a single parametric surrogate to solve the entire continuous family \textit{simultaneously} across a prescribed parameter domain $\Theta=\prod_i[\theta_i^{\min},\theta_i^{\max}] .$ The main construction of Secs.~\ref{sec:single-arch} and
\ref{sec:single-loss} is unchanged. In particular, the angular momentum is
not predicted by the network. Since it is conserved, it is fixed by the initial data as $L(\theta)=\gamma\sqrt{r_0}$ for every parameter value. However, the parameters
$\theta$ are treated as additional inputs to the network, which is therefore represented by the map
\begin{align}
(\varphi;\theta)
\;\rightarrow\;
\mathbb{F}(\varphi;\theta)
\;&\stackrel{\mathcal N}{\longmapsto}\;
\big(\mathcal N_u(\varphi;\theta),
\mathcal N_{\pr}(\varphi;\theta)\big)
\nonumber\\
&\longmapsto\;
\big(u(\varphi;\theta),p_r(\varphi;\theta)\big),
\label{eq:param-network}
\end{align}
where $\mathbb{F}$ collects the random Fourier features of Eq.~\eqref{eq:fourier-features}, the radial harmonics of Eq.~\eqref{eq:harmonic-bank} and the parameters $\theta$.
In the following, we state only what changes with respect to the previous sections.
\paragraph{Parameters as network inputs.}
The parameters $\theta$ are appended to the feature vector, so that the network sees $(\varphi,\theta)$. The output composition
\eqref{eq:u-comp}--\eqref{eq:b0} is unchanged in form, but every
quantity in it, such as $u_0(\theta)$,
$ p_{r,0}(\theta)$, $u_{\rm min}(\theta)$,
$u_{\rm max}(\theta)$, $\pr^{\max}(\theta)$, now  are evaluated per parameter point
from that orbit. The initial data are therefore enforced
exactly for \emph{every} $\theta$.
\paragraph{The radial frequency becomes a learned map.}
The harmonic bank of Sec.~\ref{sec:single-longorbit} was built on a single
number, $\omega=2\pi/\Phi_r$.
Over a family this number becomes a function of the parameters,
$\omega(\theta)=2\pi/\Phi_r(\theta)$, and the harmonic features
$\mathbb{F}(\varphi,\theta)$ are now evaluated
per collocation point. We represent $\omega(\theta)$
by a small auxiliary network, $4\to32\to32\to1$, where its four inputs are the components of
$\theta$, and its output is the corresponding radial frequency. The auxiliary
network is first fitted to the frequencies measured from a set of anchor
orbits,
$
\omega_k^{\rm ref}
=
{2\pi}/{\Phi_r(\theta_k)}.
$
The corresponding supervised loss is
\begin{equation}
\ell_\omega
=
\frac{1}{N_{\rm anc}}
\sum_{k=1}^{N_{\rm anc}}
\left[
\omega(\theta_k)-\omega_k^{\rm ref}
\right]^2 ,
\label{eq:omega-loss}
\end{equation}
where $N_{\rm anc}$ is the number of anchor orbits. After this pre-fit, the auxiliary network is used to provide
$\omega(\theta)$ during the training of the main surrogate. The pre-fit is useful because, as we explained in the previous section, an error in the radial frequency produces a
phase error that grows with the integration interval. Fitting $\omega(\theta)$ to the anchor frequencies therefore
provides the network with the correct radial scale before it is used to
represent the orbital dynamics.

After the pre-fit, the main network is trained with the physics loss introduced
for the single-orbit problem, with two differences: the network now receives
$\theta$ as an additional input, and the harmonic features use the
parameter-dependent frequency $\omega(\theta)$. We do not use the energy term
($\lambda_E=0$): with a dense set of 2PN reference anchors it is redundant, and we
found that its gradient can compete with that of the physics residual. The resulting surrogate can therefore be evaluated at
different parameter values \textit{without retraining a separate network for each
orbit}.

\paragraph{From one anchor orbit to an anchor set.}
For the parametric problem, the data-driven loss in Eq.~\eqref{eq:anchorloss}
is evaluated using a set of $K$ \emph{anchor orbits}. Each anchor is a RK numerical
solution of the Hamiltonian at a parameter point
$\theta_k\in\Theta$. The points $\theta_k$ are selected using a Latin
hypercube design and include the boundaries of the parameter domain. We check
explicitly that every selected point corresponds to a bound orbit ($E_{N,0}<0$).

The reason for using multiple anchors is different from the single-orbit case.
For one orbit, the initial condition fixes the phase at $\varphi=0$, and the
anchor mainly helps the optimizer reach the correct solution. For a parametric
family, the phase has to be fixed for every value of $\theta$. Between anchor
points, the network has to interpolate this phase information from the nearby
anchors. If the anchors are too sparse, the phase in those regions is
determined mainly by the smoothness of the network rather than by reference
data.
For this reason, we expect the distribution of the anchors across $\Theta$ to
be more important than the accuracy of any individual anchor orbit. In the
following, we therefore study how the accuracy of the surrogate changes with
the density of anchor points in the parameter domain (Sec.~\ref{sec:box}).
\paragraph{The weight $\alpha$ depends on the PN order.}
Whether the anchor should be removed during training depends on whether the
anchor orbits satisfy the same equations as those enforced by the physics loss.
In the single-orbit transfer of Sec.~\ref{sec:single-train} they do not (the
anchor is a 1PN orbit), so $\alpha$ is reduced to zero. Here the anchors are 2PN
solutions of the same equations enforced during training, so the anchor loss and
the physics loss share the same solution and there is no reason to remove the
anchor. We therefore keep $\alpha=1$ throughout training, and only $\beta$
follows the curriculum, increasing from zero to one.\footnote{Lowering the anchor
weight to $\alpha=0.1$ after the curriculum reduces the physics residual by a factor
of four---the intended effect---but \emph{raises} the trajectory error: the fraction of
the benchmark below $10^{-3}$ falls to $68.8\%$.}

\paragraph{The loss becomes a five-dimensional integral.}
The physics loss of Eq.~\eqref{eq:pdeloss} is an integral of the
squared residual over the training domain. For a single orbit that domain is
the interval $[0,\varphi_{\max}]$; for the parametric surrogate it is the
product $\mathcal{D}=[0,\varphi_{\max}]\times\Theta$,
\begin{equation}
\ell_{\rm PDE}=\frac{1}{|\mathcal{D}|}
\int_{\Theta}\!\!\int_{0}^{\varphi_{\max}}
\Big[R_u^2+R_{\pr}^2\Big]\,\mathrm{d}\varphi\,\mathrm{d}^4\theta ,
\label{eq:pdeloss-param}
\end{equation}
a five-dimensional integral that must be re-evaluated at every gradient step.

Evaluating it on a tensor-product grid is not viable. A grid with $n$ points
per axis costs $n^5$ evaluations, so even  $n=10$ requires $10^5$
points per step while resolving nothing in $\varphi$, where the orbit
completes about nine revolutions over the domain. We therefore estimate
Eq.~\eqref{eq:pdeloss-param} by Monte-Carlo integration, drawing $N_c=8000$ points
$(\varphi_j,\theta^{(j)})$ uniformly at random from $\mathcal{D}$ and averaging,
\begin{equation}
\ell_{\rm PDE}\simeq\frac{1}{N_c}\sum_{j=1}^{N_c}
\Big[R_u(\varphi_j;\theta^{(j)})^2+R_{\pr}(\varphi_j;\theta^{(j)})^2\Big].
\label{eq:pdeloss-mc}
\end{equation}
Here, $\theta^{(j)} \in \Theta$ is the $j$-th  point chosen with Monte Carlo in the 4-dimensional space $\Theta$. The statistical error of such
an estimate falls as $N_c^{-1/2}$ \emph{independently of dimension}, which is
what makes the cost of the parametric problem comparable to that of the
single-orbit one despite the four extra integration variables.

\subsection{Results } 
\label{sec:results} 
We evaluate the surrogate against independent RK integrations of the same Hamiltonian and at the same parameter point $\theta$. Unless a parameter point is part of the anchor set used for training, its reference trajectory is never provided to the network.

\subsubsection{Parameter box and training set } 
\label{sec:box} 
Recalling that $\theta=(\nu,r_0,p_{r,0},\gamma)$, we consider the four-dimensional parameter domain 
\begin{equation}
\Theta = [0.001,\,0.25]\times[0.5,\,2]\times[-0.35,\,0.35]\times[0.6,\,1.2].
\label{eq:box}
\end{equation}
This range includes both initially infalling and outgoing orbits and extends into the near-test-mass regime, down to $\nu=10^{-3}$. The corresponding Newtonian eccentricities span approximately $e_0\in[0.02,0.74]$.

The angular domain is common to all orbits. Each reference orbit is integrated for ten Newtonian periods, which yields an angular span $\varphi_i$ that varies with $\theta$. We set $\varphi_{\max}=\min_i\varphi_i$ over the anchor set, resample all anchors onto $[0,\varphi_{\max}]$, and draw the collocation points of Eq.~\eqref{eq:pdeloss-mc} on this same interval for every $\theta$. This gives $\varphi_{\max}=55.5$~rad ($\simeq8.8$ revolutions) for $K=5\times10^{3}$ and $2\times10^{4}$, and $57.9$~rad ($\simeq9.2$ revolutions) for $K=500$. Test orbits are evaluated on $[0,\min(\varphi_i,\varphi_{\max})]$. Because of the periapsis advance, the window contains a $\theta$-dependent number of radial periods, $N=\varphi_{\max}/\Phi_r(\theta)$, which is always below nine. (The $K=500$ rows of Table~\ref{tab:supervised} thus refer to a slightly longer window.)

Two aspects of the training set are important: the number of anchor
orbits and how often each anchor is sampled during training. Increasing the
number of anchors improves the coverage of the parameter space, but, for a
fixed minibatch size and training budget, each anchor is then sampled less
often. We therefore scale the anchor count and the supervised
minibatch together.
We use $K=5\times10^{2}$ anchor orbits with a  minibatch of $8000$ points, $K=5\times10^{3}$ and $2\times10^{4}$ anchors with a minibatch of $16000$ points, and
a trunk of width $384$ and depth $6$ for all of them.

\subsubsection{Accuracy over the parameter box} \label{sec:benchmark} 
We evaluate the surrogate on a benchmark of $10^{5}$ orbits drawn uniformly from the parameter box $\Theta$. With $K=2\times10^{4}$ anchors it achieves a median $\Ltwo$ error of $3.7\times10^{-4}$, with $94.9\%$ of the orbits below $10^{-3}$. A purely supervised network with the same architecture and anchors reaches $93.9\%$, so at this anchor density the physics loss contributes little to the headline accuracy; its role is largest at lower densities, $K\le5\times10^{3}$ (Sec.~\ref{sec:supervised}).

We also examine how the error depends on the distance from the training data. The $10^{5}$ test points are divided into quartiles according to their distance from the nearest anchor in normalized parameter space. Each component of $\theta$ is mapped affinely using its range in
Eq.~\eqref{eq:box}, and distances are Euclidean in this normalized space, so
that all four parameters contribute comparably regardless of their physical
units. The quartile closest to an anchor has a median $\Ltwo$  error of $3.6\times10^{-4}$,
with $96\%$ of the orbits below $10^{-3}$. For the quartile farthest from the
anchors, the corresponding values are $4.0\times10^{-4}$ and $93\%$. Thus, even
across the full range of distances from the training set, the change in median $\Ltwo$ 
error is only about $11\%$. This small variation shows that the accuracy is nearly uniform over the box at this anchor density. It is not by itself evidence of physics-driven interpolation: the purely supervised model shows a similar degradation with distance at the same $K$ (Sec.~\ref{sec:supervised}), so at $K=2\times10^{4}$ the uniformity is largely a consequence of the dense anchor set.

\begin{figure*}[t]
    \centering
    \includegraphics[width=0.9\textwidth]{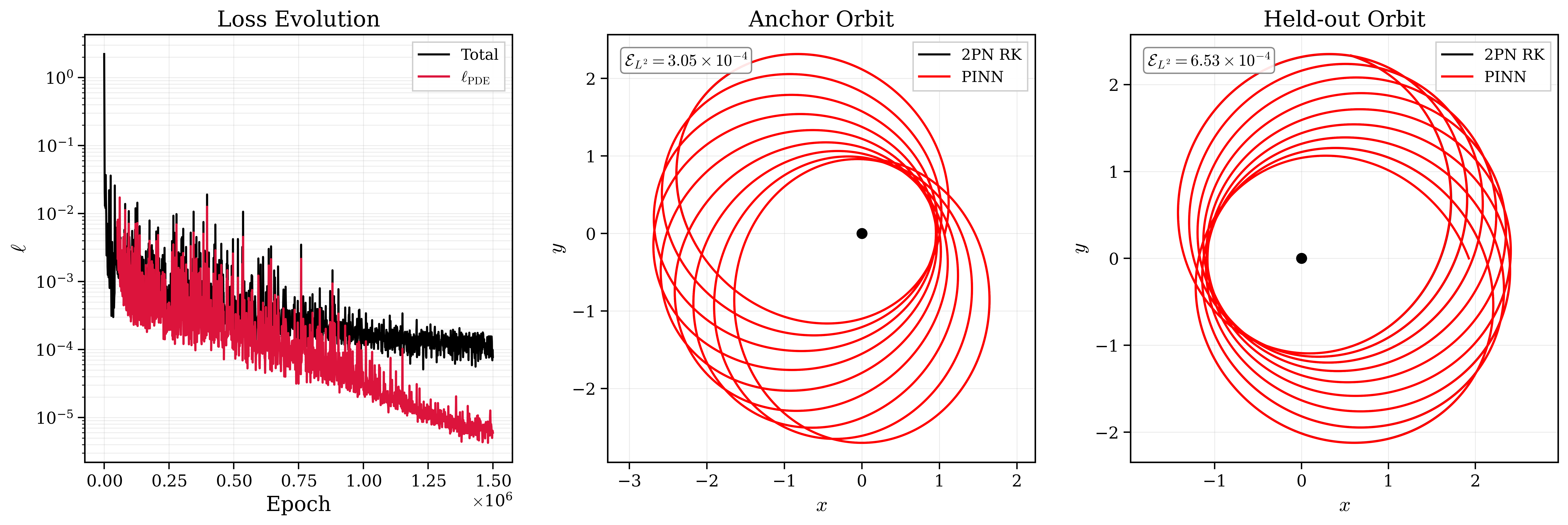}\hfill
    \caption{
Training behavior and representative predictions of the parametric surrogate after $1.5\times10^{6}$ iterations, over the common domain $\varphi\in[0,55.5]$~rad ($\simeq8.8$ revolutions).
The left panel shows the evolution of the total loss and of the physics contribution $\ell_{\rm PDE}$.
The center panel compares the PINN prediction with the 2PN RK  solution for the anchor configuration
$\theta=(0.07,1.07,0.3,1.19)$,
for which $\mathcal{E}_{L^2}=3.05\times10^{-4}$.
The right panel shows the corresponding comparison for the held-out configuration
$\theta=(0.1,1.14,0.23,0.79)$,
with $\mathcal{E}_{L^2}=9.13\times10^{-4}$.
}   
    \label{fig:lossparametric}
\end{figure*}

\begin{figure*}[t]
    \centering
    \includegraphics[width=0.95\textwidth]{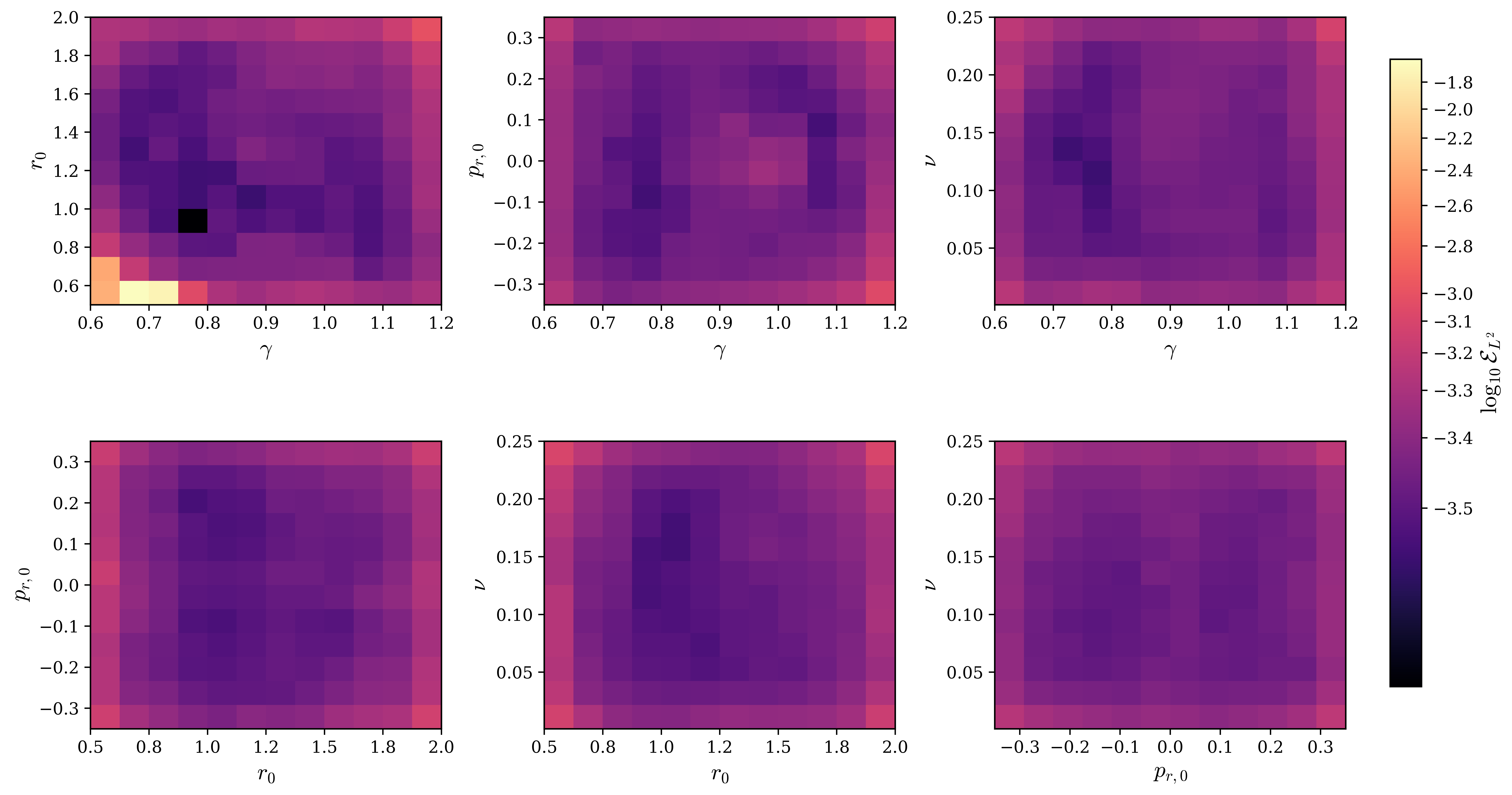}
   \caption{
Pairwise heat maps of $\mathcal{E}_{L^2}$ across the four-dimensional parameter space, evaluated on $10^{5}$ uniformly sampled test points.
Each panel shows the median value of $\log_{10}\mathcal{E}_{L^2}$ in $12\times12$ bins for one pair of parameters, after marginalizing over the remaining two.
Across the full test set, the median error is $\mathcal{E}_{L^2}=3.7\times10^{-4}$, with $94.9\%$ of the orbits satisfying $\mathcal{E}_{L^2}<10^{-3}$.
}
    \label{fig:6heatmaps}
\end{figure*}

This global behavior is illustrated in Figs.~\ref{fig:lossparametric} and~\ref{fig:6heatmaps}. Figure~\ref{fig:lossparametric} shows representative anchor and held-out trajectories together with the corresponding training losses, while Fig.~\ref{fig:6heatmaps} presents pairwise heat maps of the median $\Ltwo$ error across the parameter box. The error remains low throughout the domain, with only a mild degradation toward the edges of the sampled region.

\subsubsection{Where and why the surrogate loses accuracy}\label{sec:validity} The errors are not distributed uniformly across the parameter box. A first indication comes from sorting the benchmark of $10^{5}$ test orbits according to the Newtonian eccentricity. As shown in Table~\ref{ecc}, the success rate decreases as $e_0$ increases, from $98.9\%$ for $e_0<0.3$ to $77\%$ for $0.65<e_0< 0.74$.
\begin{table}[h]
\begin{ruledtabular}
\begin{tabular}{lccc}
$e_0$ & $n$ & Med $\mathcal{E}_{L^2}$ &  $<10^{-3}$ \\
\hline
$(0,0.3)$    & 32276 & $3.5\times10^{-4}$ & $98.9\%$ \\
$(0.3,0.5)$ & 42961 & $3.7\times10^{-4}$ & $97.2\%$ \\
$(0.5,0.65)$ & 23453 & $4.1\times10^{-4}$ & $86.4\%$ \\
$(0.65,0.74)$ & 1310  & $6.1\times10^{-4}$ & $77\%$ \\
\end{tabular}
\end{ruledtabular}
\caption{Benchmark of $10^{5}$ test orbits, stratified by Newtonian
eccentricity.}
\label{ecc}
\end{table}

However, eccentricity is not the most useful variable for understanding this trend. The relevant quantity for the validity of the PN expansion is the periapsis distance.
When the same $10^{5}$ test orbits are instead grouped by $\tilde{r}_{\rm p}/\tilde{r}_g$ (Table~\ref{tab:rperi}), a clearer pattern emerges. Here $\tilde r_p$ is the Newtonian estimate of Eq.~\eqref{eq:rp}; PN corrections shift the true minimum separation, most noticeably in the innermost bin, so the bin edges should be read as approximate. Averaged over eccentricity, the success rate exceeds $96\%$ in every bin above $10\,\tilde r_g$, reaching $99.1\%$ for $20<\tilde r_{\rm p}/\tilde r_g<40$, and drops to $30.5\%$ below $10\,\tilde r_g$.
\begin{table}[h]
\begin{ruledtabular}
\begin{tabular}{lccc}
$\tilde{r}_{\rm p}/\tilde{r}_g$ & $n$ & Med $\mathcal{E}_{L^2}$ &  $<10^{-3}$ \\
\hline
$(0,10)$      & 3717  & $4.5\times10^{-3}$ & $30.5\%$ \\
$(10,20)$     & 18292 & $3.7\times10^{-4}$ & $96.1\%$ \\
$(20,40)$     & 38450 & $3.4\times10^{-4}$ & $99.1\%$ \\
$(40,\infty)$ & 39541 & $3.9\times10^{-4}$ & $96.5\%$ \\
\end{tabular}
\end{ruledtabular}
\caption{Evaluation of $10^{5}$ test orbits, stratified by periapsis distance
($\tilde{r}_{\rm p}/\tilde{r}_g$).}
\label{tab:rperi}
\end{table}

The two classifications differ because eccentricity and periapsis are
correlated in this box: large eccentricity can drive the periapsis into the
strong-field region, so a one-dimensional table cannot tell which of the two
controls the error. This is particularly clear for the most difficult test cases. For example, the worst orbit in the benchmark has a Newtonian eccentricity \(e_0=0.62\) but a periapsis of approximately \(9.5\,\tilde{r}_g\). The large error is therefore not due to an exceptionally eccentric orbit, but rather to the strong-field periapsis it reaches.

The reason accuracy drops at small periapsis is that the physics loss itself gets much harder to minimize there. The residual $\ell_{\rm PDE}$ is nearly one hundred times
larger for $\tilde r_p<10\,\tilde r_g$ than for $\tilde r_p>40\,\tilde r_g$
($3.5\times10^{-5}$ versus $3.6\times10^{-7}$), while $\mathcal{E}_{L^2}$
changes by about a factor of twelve over the same range. So part of the strong-field degradation comes from the loss being harder to
optimize there.

The physics residual has a different scale at different points of the orbit. From Eq.~\eqref{eq:Ru}, the residual contains the factor $u^2$, with $u=1/r$. As $r$ decreases near periapsis, $u^2$ increases. At the same time, the derivatives of the Hamiltonian become larger in this region. As a result, an error in the predicted trajectory can produce a much larger residual near periapsis, where the errors are therefore weighted more strongly.

 \begin{table}[h]
\begin{ruledtabular}
\begin{tabular}{lcccc}
 & \multicolumn{4}{c}{$\tilde r_{\rm p}/\tilde r_g$} \\
\cline{2-5}
$e_0$ & $(0,10)$ & $(10,20)$ & $(20,40)$ & $(40,\infty)$ \\
\hline
$(0,0.3)$    & ---     & $95/80$ & $99/96$  & $99/99$ \\
$(0.3,0.5)$  & $5/7$   & $92/93$ & $99/100$ & $98/95$ \\
$(0.5,0.65)$ & $33/46$ & $99/99$ & $99/99$  & $73/62$ \\
$(0.65,0.74)$ & $30/49$ & $94/92$ & $83/89$  & $7/0$ \\
\end{tabular}
\end{ruledtabular}
\caption{Percentage of the $10^{5}$ test orbits with $\mathcal{E}_{L^2}<10^{-3}$,
stratified jointly by Newtonian eccentricity and periapsis distance, for the
physics-informed and supervised models at $K=2\times10^{4}$
(PINN\,/\,supervised). The empty cell contains no bound orbits.}
\label{tab:joint}
\end{table}

Table~\ref{tab:joint} shows that the dependence on eccentricity is largely secondary once periapsis is taken into account. For $0.5<e_0<0.65$, the fraction of successful orbits rises from $33\%$ in the innermost periapsis bin to $99\%$ for $10<\tilde r_{\rm p}/\tilde r_g<40$. Conversely, for $20<\tilde r_{\rm p}/\tilde r_g<40$, the success rate remains close to $100\%$ for all eccentricities below $0.65$. This indicates that the loss of accuracy is primarily associated with small periapsis distance, while eccentricity mainly modulates the difficulty within that regime. The two remaining low-success regions are discussed separately below.

In the strong-field region ($\tilde r_{\rm p}<10\,\tilde r_g$), the 2PN expansion loses physical validity, but since the reference is itself a 2PN solution, this does not cause the disagreement with it. The loss of accuracy is numerical: the dynamics becomes stiff near periapsis and the physics residual poorly conditioned. At $K=2\times10^{4}$ the supervised model also degrades there (Table~\ref{tab:joint}), so the difficulty is not specific to the physics loss either.

The second region consists of highly eccentric orbits with large periapsis. Here both models degrade---to $7\%$ and
$0\%$ for $e_0>0.65$ and $\tilde r_{\rm p}>40\,\tilde r_g$---so the cause is not
the loss. It lies in the output variable: the network predicts $u=1/r$, and even if its error in $u$ is small in these orbits, the conversion
$\delta r=\delta u/u^{2}$ amplifies it as the orbit widens.

In summary, the surrogate is accurate over most of the region where the 2PN
description is trustworthy. Its failures lie at close periapsis, where the
equations become stiff and the physics residual poorly conditioned, and at
distant apoapsis, where the parametrisation amplifies small errors. Both appear to be numerical rather than physical limitations. The fact that the first of these occurs in the same strong-field region where the 2PN expansion loses physical validity does not imply that one causes the other. Both arise in the same regime, but they affect different aspects of the problem: the PN breakdown limits the physical validity of the target model, while the numerical difficulty limits how accurately the surrogate reproduces that target.

\subsubsection{Is the physics loss doing the work?}
\label{sec:supervised}
The anchor orbits are used throughout training, so we test how much the physics loss contributes to the final accuracy. We compare two models with the same architecture, anchors, optimizer, and training budget of $1.5\times10^{6}$ iterations. The only difference is that the supervised model is trained without the physics term $\ell_{\rm PDE}$.
\begin{table}[b]
\begin{ruledtabular}
\begin{tabular}{lcccc}
 & \multicolumn{2}{c}{PINN} & \multicolumn{2}{c}{Supervised} \\
\cline{2-3}\cline{4-5}
$K$ & Med $\mathcal{E}_{\Ltwo}$ & $<10^{-3}$ & Med $\mathcal{E}_{\Ltwo}$ & $<10^{-3}$ \\
\hline
$500$           & $9.9\times10^{-4}$ & $50.9\%$ & $1.7\times10^{-2}$ & $0\%$ \\
$5\times10^{3}$ & $4.0\times10^{-4}$ & $92.4\%$ & $1.0\times10^{-3}$ & $46.9\%$ \\
$2\times10^{4}$ & $3.7\times10^{-4}$ & $94.9\%$ & $4.8\times10^{-4}$ & $93.9\%$ \\
\end{tabular}
\end{ruledtabular}
\caption{Physics-informed against purely supervised training, for three anchor
densities. All six configurations use the identical architecture,
$1.5\times10^{6}$ iterations, and the same held-out benchmark of $10^{5}$
orbits.}
\label{tab:supervised}
\end{table}

The effect of the physics loss depends strongly on the number of anchors
(see Table~\ref{tab:supervised}), with the transition from
physics-dominated to data-dominated performance occurring between
$K=5\times10^{3}$ and $K=2\times10^{4}$. With $K=500$, the physics-informed model has a median $\Ltwo$ error of $9.9\times10^{-4}$, compared with $1.7\times10^{-2}$ for the supervised model. The fraction of test orbits with $\mathcal{E}_{\Ltwo}<10^{-3}$ is $50.9\%$ for the physics-informed model and $0\%$ for the supervised one. Thus, with a sparse anchor set, the physics loss provides a substantial improvement.
With $K=2\times10^{4}$, the difference is much smaller. The median $\Ltwo$ errors are $3.7\times10^{-4}$ and $4.8\times10^{-4}$, and the fractions below $10^{-3}$ are $94.9\%$ and $93.9\%$, respectively. With this many anchors, the supervised model already has enough reference orbits to reproduce the solution accurately over most of the parameter space. The improvement from the physics loss is therefore limited. 

At the intermediate density $K=5\times10^{3}$, the physics loss already provides a substantial advantage. The PINN reaches a median $\mathcal{E}_{\Ltwo}=4.0\times10^{-4}$, with $92.4\%$ of the orbits below $10^{-3}$, compared with $46.9\%$ for the purely supervised model. Increasing the anchor set to $2\times10^{4}$ changes the PINN success rate only marginally, from $92.4\%$ to $94.9\%$, while the supervised model improves from $46.9\%$ to $93.9\%$. In this comparison the physics loss therefore saves up to roughly a factor of four in anchor count: the PINN at $K=5\times10^{3}$ ($92.4\%$) is comparable to the supervised model at $K=2\times10^{4}$ ($93.9\%$). The supervised $K$ needed to match $92.4\%$ lies between $5\times10^{3}$ and $2\times10^{4}$ and was not determined.

The orbit-by-orbit comparison shows that for $K=2\times10^{4}$ the
physics-informed model is more accurate on $82.0\%$ of the test orbits, with a
median ratio of the supervised to the PINN error of $1.30$, i.e.\ the median error of the PINN is $1.3$ times lower. For $K=500$, the difference is much larger: the physics-informed model is more accurate on $98.4\%$ of the orbits, with a median supervised/PINN error ratio of $17.4$. No orbit satisfies the threshold with the supervised model while failing it with the physics-informed model.

We further compare the dependence of the error on the distance from the training anchors in parameter space by separating the benchmark into the nearest and farthest quartiles. For $K=500$, the physics-informed and supervised models show a similar degradation with distance: the ratio between the median error in the farthest and nearest quartiles is $1.38$ for the physics-informed model and $1.31$ for the supervised baseline. At this density the anchor set is too sparse to constrain even the local neighbourhoods well, so the physics residual improves the solution across the parameter space without a clear advantage in the regions between anchors.
At $K=5\times10^{3}$, the difference becomes more evident. For the physics-informed model, the fraction of orbits with error below $10^{-3}$ decreases from $94\%$ in the nearest quartile to $89\%$ in the farthest one. For the supervised model, the same fraction drops from $52\%$ to $37\%$. At this intermediate anchor density, the training data constrain the regions close to the anchors, while the physics residual improves the interpolation in the regions farther from them.
For $K=2\times10^{4}$, the two models again show a similar degradation between the nearest and farthest quartiles. In this regime the anchor set is sufficiently dense that supervised interpolation alone is already effective, and the additional benefit of the physics residual becomes small.

Physics-informed training takes about six times longer than supervised training in our setup, so its benefit is small when a large number of reference orbits is available. For the PN problem, generating $2\times10^{4}$ reference orbits takes less than two hours at about $0.25\mathrm{s}$ per orbit (at the tolerances of Sec.~\ref{sec:single}), making the dense-anchor regime feasible. For the numerical-relativity problem considered as the long-term target, however, a single reference orbit costs about $\Ord(10^{5})$ CPU-hours. Generating $K\sim10^{4}$ reference orbits is therefore impractical, making the sparse-anchor regime the relevant one.

\section{Dissipative dynamics at 2.5PN order}
\label{sec:dissipative}
The conservative dynamics of Sec.~\ref{sec:phase} present a phase-determination challenge for physics-only training. Because Eq.~\eqref{eq:angle-system} is autonomous and its solutions are quasi-periodic, the physics loss cannot distinguish the orbit from its phase translates, and the phase must be supplied by the initial data and, where needed, by supervised data.
Radiation reaction changes the nature of the difficulty. The dissipative system
is still autonomous in $\varphi$, so its residual remains exactly translation
invariant. We find that the $2.5$PN inspiral can be recovered, with errors below about $10^{-3}$ for $e_0N_{\rm rev}\lesssim3.7$ and degrading beyond (Sec.~\ref{sec:diss-results}), where $N_{\rm rev}=\varphi_{\max}/2\pi$ is the number of revolutions covered, which exceeds the number of radial periods because of the periapsis advance, without any $2.5$PN reference trajectory in the loss, starting from a warm start on the
conservative $2$PN orbit, which provides the initial orbital shape and phase,
and using only the $2$PN Hamiltonian, the analytic radiation-reaction force, and
an energy-balance residual.
There is, however, an important numerical issue: the radiation-reaction force
is much smaller than the conservative force, and its effect is therefore not
well resolved by a standard pointwise residual. The formulation and loss have
to be modified accordingly. We describe these changes below.

\subsection{Dissipative dynamics in the angle domain }
\label{sec:diss-formulation}
At $2.5$PN order, we combine the conservative Hamiltonian truncated at
2PN order with the standard radiation-reaction acceleration
\cite{Lincoln:1990ji, Iyer:1995rn}

\begin{equation}
\bm a_{\rm RR}=\frac{8}{5}\frac{\nu}{c^5 r^3}
\Big[\big(3v^2+\tfrac{17}{3}\tfrac{1}{r}\big)\dot r\bm n
-\big(v^2+3\tfrac{1}{r}\big)\bm v\Big],
\label{eq:FRR}
\end{equation}
where $\bm n=\bm r/r$, $v^2=|\bm v|^2$, and
$\dot r=\bm v\cdot\bm n$. The expression is gauge dependent and is combined here with the ADM 2PN Hamiltonian; the resulting system should be regarded as a model 2.5PN system, which suffices for testing the method. 
In this case, the angular momentum is no longer conserved since the radiation-reaction force
applies a torque, so $L$ must now be evolved together with $u$ and $p_r$.
Now the system in Eq. \eqref{eq:angle-system} must also include the equation
\begin{equation}
\frac{dL}{d\varphi}=\frac{(\bm r\times\bm a_{\rm RR})_z}{\dot\varphi}.
\label{eq:diss-system}
\end{equation}
Unlike the conservative case, the solution
is not periodic: as the binary radiates energy, the orbit contracts, and the inverse radial coordinate 
$u$ undergoes a sustained secular drift to larger values. We stop each inspiral at a PN-validity cut-off, when the separation first falls below $10\,\tilde r_g$ (i.e.\ $\tilde r_g/\tilde r>0.1$), or after the prescribed number of radial periods. The $2.5$PN reference orbit is truncated at that point, and the network is trained and evaluated only on the retained portion of the trajectory; the run that reaches the cut-off early is flagged in Table~\ref{tab:diss}.

\subsection{Architecture: differences from the conservative case }
\label{sec:diss-architecture}
Each inspiral is trained as a single orbit at fixed $\theta$, using the
single-orbit architecture of Sec.~\ref{sec:single-arch}: the trunk, the random
Fourier-feature banks, the exact enforcement of the initial conditions through
the gate $g(\varphi)$, and the orbit-adapted output range for $u$ are unchanged.
The dissipative problem requires three modifications of the network and its
loss, described below, together with a warm start from the conservative orbit.

\paragraph*{Angular momentum becomes a trainable output.}
In the conservative problem, $L$ is a constant fixed by the initial data, and its conservation is imposed exactly by the network's construction. This is no longer true once radiation reaction is included, as the angular momentum decays. We therefore promote $L$ to a third network output, $L(\varphi)$. To impose its initial value $L_0$ exactly while guaranteeing that it  remains strictly positive during orbital decay, we apply an initial-data gate via the ansatz $L(\varphi) = L_0 \frac{\text{softplus}(1 + g(\varphi)\mathcal{N}_L(\varphi))}{\text{softplus}(1)}$, where $g(0)=0$. Its subsequent evolution is then determined by the residual of Eq.~\eqref{eq:diss-system}.

\paragraph*{The radial-harmonic bank is disabled.}
The most consequential architectural change is the removal of the
radial-harmonic features introduced for the long conservative orbits in
Sec.~\ref{sec:single-longorbit}. Those features presuppose a fixed radial
frequency and a fixed oscillation amplitude, represented by harmonics
$\{\cos(n\omega\varphi),\sin(n\omega\varphi)\}$ with constant coefficients.
This is exactly right for a closed, precessing conservative orbit and wrong
for an inspiral, whose amplitude decreases from one radial period to the
next: a fixed-amplitude basis cannot represent a decaying oscillation, and in
practice the network minimises the residual by collapsing towards a
near-constant radius. We therefore set $N_h=0$ and retain only the random
Fourier-feature banks of Eq.~\eqref{eq:fourier-features}, which carry the
secular evolution directly.

\paragraph*{The energy anchor is removed.}
In the conservative single-orbit case, the energy penalty
$\ell_E$ provides an additional constraint during
training. Because orbital energy is no longer conserved in the dissipative case, this penalty is omitted ($\lambda_E=0$, and no $\beta$-independent energy term).
\paragraph*{Warm start.}
We initialize the dissipative training using a supervised fit to the
conservative $2$PN orbit. This does not introduce radiation-reaction
information into the training objective: the reference is a solution of the
conservative dynamics only. Its role is to provide a reasonable initial
representation of the orbital shape and phase before the dissipative
equations are introduced.

\subsection{The obstruction: radiation reaction is not resolved by a pointwise residual}
\label{sec:diss-obstruction}
A direct extension of the physics-informed loss used for the conservative
problem is not sufficient for the inspiral. The reason is the relative size
of the radiation-reaction force. Along the inspirals considered here,
$|\bm a_{\rm RR}|/|\bm a_{\rm cons}|$ is between
$2.6\times10^{-5}$ at the beginning and $2.6\times10^{-4}$ at the end of
the evolution, with a maximum of $2.1\times10^{-3}$ near periapsis.
Therefore, the reactive contribution is much smaller than the conservative
one at most points of the orbit.
This creates a problem for a loss based only on the pointwise equations of
motion. In our runs, the root-mean-square PDE residual can be reduced to ${\sim}3\times10^{-3}$ (i.e.\ $\ell_{\rm PDE}\sim10^{-5}$) without resolving the radiation-reaction term. In other words,  the conservative dynamics is already satisfied to good
accuracy, while the dissipative contribution remains below the scale of the
residual. The optimizer can therefore reduce the loss without learning the
inspiral:  the loss is already ${\sim}10^{-5}$, but the relative
orbit error remains close to $2\times10^{-1}$ and the resulting trajectory is
a bounded, non-decaying rosette rather than an inspiral. 

With the pointwise loss alone, adding causal
weighting~\cite{WangPerdikaris2022}, curriculum scheduling, or domain
marching does not remove this behavior. These methods change the way in
which the pointwise residual is optimized, but they do not add information
about the small secular effect that distinguishes the inspiral from the
conservative solution. As we show below, domain marching becomes effective
only once this information is supplied by the energy-balance term of
Sec.~\ref{sec:diss-loss}.
The relevant information is therefore not only the instantaneous size of
the radiation-reaction force, but its accumulated effect over many
orbital cycles. Although the reactive force is only of order
$3\times10^{-4}$ of the conservative force over most of the orbit, its
effect accumulates coherently and removes approximately $15\%$ of the
binding energy over the revolutions considered here. We therefore
supplement the pointwise equations with an energy-balance residual, as we are going to see in the following section.
A similar conclusion was reached in Ref.~\cite{GomesDaSilva2026} for learned
flow maps of quasi-circular $2.5$PN inspirals: pointwise training objectives
are blind to the secular drift, and long-time accuracy is controlled by the
channel fixed by energy--angular-momentum balance, which is imposed there on
the map output after training. Here, the balance law enters directly as a
residual in the training loss, and the inspirals considered are eccentric.

\subsection{The secular balance-law loss }
\label{sec:diss-loss}
The energy balance provides a direct way of constraining the cumulative
effect of radiation reaction. The energy
change along the dissipative trajectory satisfies
\begin{equation}
\frac{d\hat H}{d\varphi}
=\frac{\bm a_{\rm RR}\cdot\bm v}{\dot\varphi},
\label{eq:balance}
\end{equation}
which gives the additional residual
\begin{equation}
R_E^{(2.5{\rm PN})}=\frac{1}{\Lambda}\left[
\frac{d}{d\varphi}\hat H\big(u,\pr,L\big)
-\frac{\bm a_{\rm RR}\cdot\bm v}{\dot\varphi}\right].
\label{eq:RE}
\end{equation}
Here, $d\hat H/d\varphi$ is evaluated by automatic differentiation of the
network output, while the second term is computed directly from the
analytic radiation-reaction force. Both sides are built from the $2$PN Hamiltonian and the analytic
force~\eqref{eq:FRR}; no dissipative trajectory is involved.
The complete loss is then
\begin{equation}
\ell=\alpha\ell_{\rm anchor}^{(2{\rm PN})}
+\beta\Big(\ell_{\rm PDE}+w_L\ell_L+w_{\rm sec}
 \ell_E^{(2.5{\rm PN})}\Big),
\label{eq:diss-loss}
\end{equation}
where $\ell_L$ is the residual associated with the angular-momentum
equation in Eq.~\eqref{eq:diss-system} and $\ell_E^{(2.5{\rm PN})}$ the one associated with Eq. \eqref{eq:RE}. The coefficient $\alpha$ is gradually
reduced to zero during the curriculum, so that the $2$PN anchor is used only to
initialize the dissipative solution.
We use a fixed normalization for the energy residual,
$
\Lambda=1.7\times10^{-3},
$
chosen as the RMS of $\bm a_{\rm RR}\cdot\bm v/\dot\varphi$ evaluated along the conservative $2$PN warm-start orbit. This normalization makes $R^{(2.5{\rm PN})}_E$ dimensionless and
keeps the energy-balance term on a comparable scale to the pointwise PDE
residual throughout training. With this choice, the two contributions to the loss are of similar magnitude.

The pointwise residual alone admits a solution close to the conservative
$2$PN orbit, because the radiation-reaction term is small at any given point
along the trajectory. The energy balance is what distinguishes the two. To
see this, we evaluate Eq.~\eqref{eq:RE} directly on each of the two known
trajectories---the reference $2.5$PN inspiral and the conservative $2$PN
orbit---as a diagnostic, independently of training. The inspiral satisfies
the balance law by construction and gives a residual smaller by roughly three
orders of magnitude than the conservative orbit, which violates it
systematically along the entire trajectory. The mechanism is the following. The conservative flow conserves $\hat H$, so it contributes nothing to $d\hat H/d\varphi$; the only source of energy change in Eq.~\eqref{eq:balance} is the radiation-reaction term. In the pointwise equations this term is a correction of relative size $10^{-5}$--$10^{-3}$ to an $\mathcal O(1)$ force (Sec.~\ref{sec:diss-obstruction}), and is therefore below the residual scale. In $R_E^{(2.5\rm PN)}$ it is the only term left to balance, and with the normalization $\Lambda$ it enters at $\mathcal O(1)$. Moreover, the constraint prescribes the slope of $\hat H(\varphi)$ at every $\varphi$, so that, up to the integrated residual, it fixes the total energy loss over the domain. A trajectory that follows the conservative orbit violates it systematically at every angle, which is why it excludes the conservative branch.

\subsection{Input data and training protocol}
\label{sec:diss-provenance}
We do not use any dissipative trajectory to train the network. The only inputs are the $2$PN Hamiltonian, the analytic radiation-reaction force~\eqref{eq:FRR}, and the conservative $2$PN orbit used for the warm start. We do compute a $2.5$PN reference orbit, but only to evaluate the final error and for the diagnostics in Sec.~\ref{sec:diss-results}. 
This leads to an important difference from the parametric conservative case. In Sec.~\ref{sec:architecture}, keeping the anchor active ($\alpha=1$) throughout training is harmless, because the anchors solve the same equations enforced by the physics loss. Here that would be a mistake: the warm-start anchor is a non-radiating orbit, and any residual pull toward it would compete with the physics loss trying to drive the inspiral. As in the single-orbit transfer of Sec.~\ref{sec:single-train}, we therefore reduce $\alpha$ to zero during the curriculum.

The subsequent evolution
is then determined entirely by the dissipative physics loss.
For the longer inspirals, we train the network by progressively extending
the angle domain. Starting from a shorter interval, the domain is increased
in stages,
$$
[0,\varphi_1]\rightarrow[0,\varphi_2]\rightarrow\cdots
\rightarrow[0,\varphi_{\max}],
$$
using the solution from each stage to initialize the next one. This domain
marching provides a simple continuation procedure. While it does not help with
the pointwise loss alone (Sec.~\ref{sec:diss-obstruction}), once the energy-balance
term is included it gives the largest improvement in the final accuracy among
the strategies we tested.

\subsection{Results}
\label{sec:diss-results}

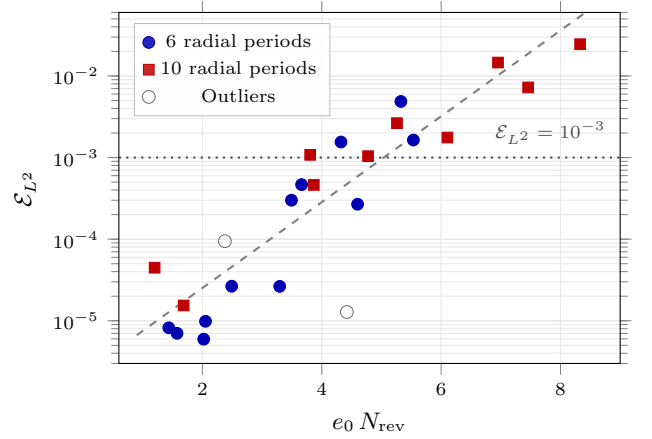
\begin{figure}[t]
\centering
\begin{tikzpicture}
\begin{axis}[
    width=0.95\columnwidth, height=0.72\columnwidth,
    xlabel={$e_0\,N_{\rm rev}$},
    ylabel={$\mathcal{E}_{\Ltwo}$},
    ymode=log,
    xmin=0.6, xmax=9.0,
    ymin=3e-6, ymax=6e-2,
    grid=both, grid style={gray!18},
    tick align=outside,
    legend style={at={(0.03,0.97)}, anchor=north west, font=\scriptsize,
                  fill=white, fill opacity=0.85, text opacity=1, draw=gray!50},
    label style={font=\small}, tick label style={font=\scriptsize},
]

\addplot[domain=0.9:8.6, samples=2, thick, gray, dashed, forget plot]
    {10^(-5.650 + 0.526*x)};

\addplot[only marks, mark=*, mark size=2.2pt, color=blue!70!black]
    coordinates {
    (1.437, 8.20e-6)   
    (1.575, 7.02e-6)   
    (2.022, 5.95e-6)   
    (2.052, 9.86e-6)   
    (2.491, 2.65e-5)   
    (3.295, 2.64e-5)   
    (3.492, 3.00e-4)   
    (3.662, 4.66e-4)   
    (4.322, 1.55e-3)   
    (4.600, 2.68e-4)   
    (5.326, 4.87e-3)   
    (5.535, 1.64e-3)   
    };
\addlegendentry{$6$ radial periods}

\addplot[only marks, mark=square*, mark size=2.0pt, color=red!75!black]
    coordinates {
    (1.199, 4.47e-5)   
    (1.686, 1.54e-5)   
    (3.806, 1.08e-3)   
    (3.866, 4.62e-4)   
    (4.776, 1.04e-3)   
    (5.260, 2.64e-3)   
    (6.105, 1.76e-3)   
    (6.953, 1.46e-2)   
    (7.458, 7.24e-3)   
    (8.329, 2.45e-2)   
    };
\addlegendentry{$10$ radial periods}

\addplot[only marks, mark=o, mark size=2.4pt, color=black!55]
    coordinates {
    (4.421, 1.28e-5)   
    (2.377, 9.45e-5)   
    };
\addlegendentry{Outliers}

\addplot[domain=0.6:9.0, samples=2, dotted, thick, black!70, forget plot]
    {1e-3};
\node[font=\scriptsize, black!70, anchor=south east] at (axis cs:8.9,1.15e-3)
    {$\mathcal{E}_{\Ltwo}=10^{-3}$};

\end{axis}
\end{tikzpicture}
\caption{Accuracy of the dissipative surrogate as a function of $e_0N_{\rm rev}$. All runs use the same training configuration. The two outliers correspond to the weak-dissipation case and the run truncated by the 2PN validity cut-off; these are excluded from the fit. The dashed line shows the least-squares relation $\log_{10}\mathcal{E}_{\Ltwo}=-5.65+0.53\,e_0N_{\rm rev}$ for the twenty-two standard configurations.
}
\label{fig:diss-scaling}
\end{figure}

\begin{figure*}[!]
    \centering
    \includegraphics[width=0.83\textwidth]{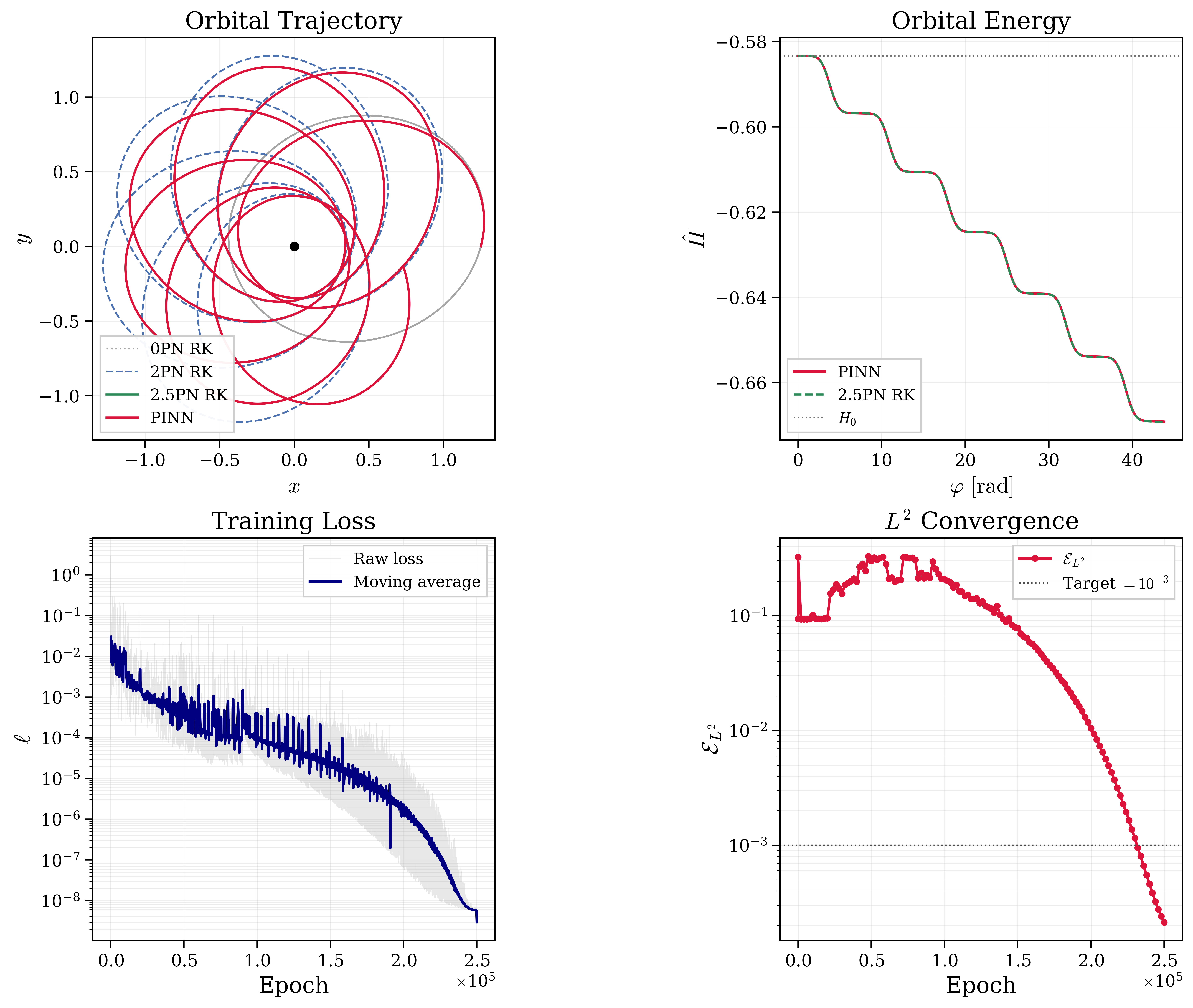}\hfill
   \caption{
Representative dissipative 2.5PN solution for
$\theta=(0.24,1.25,0.17,0.72)$, corresponding to an eccentricity $e\simeq0.5$ over six radial periods.
The upper panels show the orbital trajectory and energy evolution, while the lower panels report the training loss and the convergence of the $L^2$ error.
The final error is $\mathcal{E}_{L^2}=2.14\times10^{-4}$, below the target accuracy of $10^{-3}$.
}
    \label{fig:lossdissipative}
\end{figure*}

We assess the dissipative surrogate using the relative $\Ltwo$ error and the orbital energy $\hat H(\varphi)$. Although $\hat H(\varphi)$ is not fitted directly, its slope is constrained by Eq.~\eqref{eq:RE}; the energy curve therefore tests the integrated effect of the constraint and the consistency of the trajectory, not an independent prediction.

For a representative configuration,
$
\theta=(0.24,1.25,0.17,0.72),
$
evolved over six radial periods, Fig.~\ref{fig:lossdissipative} shows the corresponding orbital trajectory, energy evolution, training loss, and convergence of the $\Ltwo$ error. The surrogate reaches
\begin{equation}
\mathcal{E}_{\Ltwo}=2.14\times10^{-4},
\qquad
\mathcal{E}_{\Ltwo}^{2-2.5}=5.1\times10^{-2},
\label{eq:diss-reference}
\end{equation}
where, $\mathcal{E}_{\Ltwo}^{2-2.5}$ is the separation between the conservative $2$PN and dissipative $2.5$PN trajectories, measured with Eq.~\eqref{eq:L2_error}. The network therefore recovers
$
1-{\mathcal{E}_{\Ltwo}}/
{\mathcal{E}_{\Ltwo}^{2-2.5}}
=99.4\%
$
of the dissipative correction. Over the same interval, the predicted orbital energy decreases by $14.7\%$ relative to $|\hat H_0|$.

The full sample comprises twenty-four configurations evolved over six or ten radial periods; Table~\ref{tab:diss} lists a representative subset of nine, chosen to span the range of the sample and to illustrate the points made below, while Fig.~\ref{fig:diss-scaling} plots all twenty-four
dissipative-sector configurations. The sample spans initial eccentricities $e_0\in[0.1,0.75]$, and periapsis distances $\tilde r_{\rm p}/\tilde r_g\in[9,37.4]$. All runs use the same training setup; only the initial conditions and evolution length are changed. Two configurations are discussed separately below because they do not provide standard tests of the dissipative sector. Across the remaining twenty-two cases, the median $\Ltwo$ error is $4.6\times10^{-4}$, twelve runs satisfy $\mathcal{E}_{\Ltwo}<10^{-3}$, and the median recovered fraction of the $2$PN--$2.5$PN correction is $99.2\%$. All configurations show a net decrease in orbital energy. Several of them have $\tilde r_{\rm p}<20\,\tilde r_g$ and thus lie outside the PN-reliable label of Sec.~\ref{sec:formulation}; they test the method on the 2.5PN mathematical system.

The dominant trend in the error is associated with the combination $e_0N_{\rm rev}$. Its correlation with $\log_{10}\mathcal{E}_{\Ltwo}$ is $\rho=0.93$, compared with $0.75$ for $e_0$, $0.50$ for $N_{\rm rev}$, and $0.34$ for $\mathcal{E}_{\Ltwo}^{2-2.5}$. A least-squares fit over the twenty-two standard configurations gives
\begin{equation}
\log_{10}\mathcal{E}_{\Ltwo}
\simeq
-5.65+0.53e_0N_{\rm rev},
\label{eq:diss-scaling}
\end{equation}
with a coefficient of determination $R^2=0.87$ (see Fig. \ref{fig:diss-scaling}). Adding $\log_{10}\mathcal{E}_{\Ltwo}^{2-2.5}$ as an additional regressor increases $R^2$ only to $0.93$.

The combination $e_0N_{\rm rev}$ captures two aspects of the problem: the sharpness of each periapsis passage and the number of times that structure must be represented. At $e_0=0.75$, the orbital radius varies by about a factor of seven over one radial cycle, producing a pronounced maximum in $u=1/r$ near periapsis. Since $|\bm a_{\rm RR}|\propto r^{-3}$, the radiation-reaction force also changes strongly along the orbit. As the number of revolutions increases, these localized features must be reproduced repeatedly over a fixed representation. Indeed, the number of Fourier modes required to capture $99\%$ of the signal's spectral power rises from $7$ for near-circular cases to $50$ for the most demanding ones, exceeding the $N_{\rm L}+N_{\rm H}=26$ random frequencies of the fixed Fourier embedding. 


The training schedule also affects the final accuracy. In particular, it is important to leave a sufficient fraction of the optimization budget after the full angular domain has been introduced. All runs reported here use the same fixed iteration budget, and training stops while the error is still decreasing. The values in Table~\ref{tab:diss} should therefore be interpreted as the accuracy obtained at fixed computational cost.
The two configurations highlighted in Table \ref{tab:diss} require separate comment. In one case, the $2$PN and $2.5$PN trajectories are separated by only $1.4\times10^{-3}$ and the total energy change is below $0.3\%$. This case therefore provides only a weak test of the dissipative correction. In another configuration, the post-Newtonian validity cut-off is reached before six radial periods are completed, so the reported error is evaluated only over the portion of the trajectory that remains within the valid regime.

\begin{table*}[!]
\begin{ruledtabular}
\begin{tabular}{ccccccc}
$e_0$ &
$N_{\rm rev}$ &
$\tilde r_{\rm p}/\tilde r_g$ &
$\mathcal{E}_{\Ltwo}$ &
$\mathcal{E}_{\Ltwo}^{2-2.5}$ &
$1-\mathcal{E}_{\Ltwo}/\mathcal{E}_{\Ltwo}^{2-2.5}\,(\%)$ &
$\Delta\hat H\,(\%)$ \\
\hline
 
\multicolumn{7}{c}{\emph{6 radial periods}}\\
\hline
 
$0.23$ & $6.36$ & $37.4$ & $8.2\times10^{-6}$ & $5.5\times10^{-3}$ & $99.9\%$ & $-1\%$ \\
$0.28$ & $7.38$ & $15$ & $9.9\times10^{-6}$ & $6.6\times10^{-2}$ & $100\%$ & $-14.4\%$ \\
$0.5$ & $6.97$ & $13.9$ & $2.1\times10^{-4}$ & $5.1\times10^{-2}$ & $99.4\%$ & $-14.7\%$ \\
$0.7$ & $6.6$ & $15.2$ & $2.7\times10^{-4}$ & $3.7\times10^{-2}$ & $99.3\%$ & $-9.9\%$ \\
$0.58$ & $4.12$\footnote{Post-Newtonian validity cut-off reached before the sixth radial period; error quoted over the shortened domain.} & $9$ & $9.5\times10^{-5}$ & $5.6\times10^{-2}$ & $99.8\%$ & $-24.6\%$ \\
$0.57$ & $7.7$ & $29.5$ & $1.3\times10^{-5}$ & $1.4\times10^{-3}$\footnote{Weak-dissipation regime: the radius increases over the window and the energy changes by only $0.3\%$. Not a test of the dissipative sector.} & $99.1\%$ & $-0.3\%$ \\
 
\hline
\multicolumn{7}{c}{\emph{10 radial periods}}\\
\hline
 
$0.1$ & $12.23$ & $25.4$ & $4.5\times10^{-5}$ & $5.7\times10^{-2}$ & $99.9\%$ & $-10.7\%$ \\
$0.44$ & $10.93$ & $21.1$ & $1\times10^{-3}$ & $2.9\times10^{-2}$ & $96.4\%$ & $-6.2\%$ \\
$0.7$ & $11.95$ & $15.2$ & $2.5\times10^{-2}$ & $6.5\times10^{-2}$ & $62.1\%$ & $-17.4\%$ \\
\end{tabular}
\end{ruledtabular}
\caption{$2.5$PN inspirals trained without dissipative reference data (conservative $2$PN warm start): a representative subset of nine
configurations out of the full sample of twenty-four described in the text.  $\mathcal{E}_{\Ltwo}$
is the relative phase-space error with respect to the $2.5$PN RK solution,
while $\mathcal{E}_{\Ltwo}^{2-2.5}$ denotes the corresponding
$2$PN--$2.5$PN separation. The captured fraction is
$1-\mathcal{E}_{\Ltwo}/\mathcal{E}_{\Ltwo}^{2-2.5}$. $\Delta\hat H$ is the
total energy change relative to $|\hat H_0|$.}
\label{tab:diss}
\end{table*}

\section{Discussion and outlook}
\label{sec:discussion}
 
In this work we introduced \texttt{GravINNs}, a gravity-informed neural-network
framework for post-Newtonian binary dynamics. We considered conservative
single-orbit solutions, long integrations, a four-dimensional parametric
surrogate, and dissipative inspirals including the leading $2.5$PN
radiation-reaction contribution. The approach works when the physics is
embedded where it belongs: conservation laws and initial data in the
architecture, radial periodicity in the features, and the equations of motion
in the loss. Over the full
parameter box, the surrogate achieves a median
$\Ltwo$ error of $3.7\times10^{-4}$, with $94.9\%$ of the orbits below
$10^{-3}$ ($K=2\times10^{4}$ anchors; a purely supervised network reaches
$93.9\%$ at this density; the physics loss matters most for
$K\le5\times10^{3}$, Sec.~\ref{sec:supervised}). Restricted to the
region where the post-Newtonian expansion is itself reliable,
$\tilde r_p>\tilde r_{\rm bd}=20\,\tilde r_g$, this fraction rises to $98\%$.
 
The different regimes exposed two limitations of physics-informed training,
and they are opposites. In the conservative problem the autonomous
angle-domain residual is invariant under rigid rotations
(Sec.~\ref{sec:phase}), so the residual alone fixes the orbit only up to a phase, and for quasi-periodic
orbits the exact initial condition leaves an almost flat direction in the loss:
the residual is \emph{nearly degenerate}, and reweighting it alone does not remove it. The main effect is on reliability rather than accuracy. For near-circular orbits, physics-only training converges to the wrong phase in $38\%$ of random initializations, even though the residuals are comparable to those of successful runs. The same phase degeneracy also motivates the use of a representation adapted to the radial periodicity in long integrations.

Under radiation reaction the system remains autonomous, and the difficulty is of a
different kind: the radiation-reaction term is three to four orders of magnitude
below the conservative dynamics pointwise, so the residual is no longer
\emph{nearly degenerate} but \emph{ill-conditioned}, and the pointwise loss is
minimized by a non-decaying rosette.

The solutions are equally opposite. The conservative
case can need additional phase information for reliable optimization, supplied by sparse data or by transfer from a
lower PN order; the dissipative case requires no dissipative trajectory data (it still uses a warm start from the conservative orbit), but it does require the right global constraint: an energy-balance relation that makes the secular radiation-reaction effect visible to the loss. The recipe in both is the same: determine what the residual cannot
see, and supply it explicitly.
 These results also help place physics-informed surrogates in relation to
standard numerical methods. For a single post-Newtonian orbit, direct
numerical integration remains faster and more accurate. The advantage appears
when many solutions are required across a parameter space: once trained, a
single network approximates an entire family of trajectories without a new
integration for each parameter choice, and the physics loss reduces the
reference data needed to cover that space. Its value is concentrated where
reference solutions are expensive, which is the regime of the long-term
target, since a single numerical-relativity orbit costs $\Ord(10^{5})$
CPU-hours. The surrogate is accurate throughout the region where the $2$PN description is
trustworthy, and its failures concentrate at close periapsis passages and at
the distant apoapsis of strongly eccentric orbits (Sec.~\ref{sec:validity}).
The first coincides with the region where the expansion itself becomes questionable, but since the reference is a 2PN solution, the failure is numerical: it is where the physics residual is most poorly conditioned, and the supervised model also degrades there.

The post-Newtonian two-body problem therefore serves not only as an
application, but also as a controlled setting in which the main challenges of
physics-informed learning can be isolated and tested against accurate
numerical solutions. The route ahead is to scale these principles to higher-dimensional parameter spaces (spins, higher PN orders) and to operator-valued
targets~\cite{Li2021, Lu2021}; to carry the dissipative surrogate from orbital dynamics to radiated multipoles; and to extend the surrogate into the strong field, both by substituting
effective-one-body dynamics for the truncated Hamiltonian and by improving the
conditioning of the physics residual near periapsis
(Sec.~\ref{sec:validity}).
This is a natural starting point for moving toward the
field equations of general relativity, where constraint preservation, gauge
freedom, and a much larger solution space become unavoidable. There the
analogue of the phase degeneracy is already familiar: a small
evolution-equation residual does not guarantee that the Hamiltonian and
momentum constraints are satisfied~\cite{Gundlach2005}. As the system becomes more complex, the number of weakly constrained directions is expected to grow, making transfer strategies increasingly important for realistic targets. We are currently pursuing this extension toward
physics-informed solutions of the Einstein
equations, with \texttt{GravINNs} providing the
controlled foundation for this broader program.

\begin{acknowledgments}
We thank Romina Ballesteros, Miriam Patricolo, Francesco~Sacco and Matteo~Zatti for interesting remarks and discussions. JM thanks Paola Castillo for her constant and unconditional support.
The work of GB and JM has been supported in part by the MCI, AEI,
FEDER (UE) grant PID2024-155685NB-C21 ``Gravity, Supergravity and
Superstrings'' (GRASS) and ``IFT Centro de Excelencia Severo Ochoa''
CEX2020-001007-S. The work of GB has been supported by the fellowship
CEX2020-001007-S-20-5. AI tools were used to assist with manuscript editing and code development. All AI-assisted content was reviewed and verified by the authors.
\end{acknowledgments}


\begin{thebibliography}{99}

\bibitem{Abbott2016}
B.~P.~Abbott \emph{et al.} (LIGO Scientific and Virgo Collaborations),
``Observation of Gravitational Waves from a Binary Black Hole Merger,''
Phys.\ Rev.\ Lett.\ \textbf{116}, 061102 (2016).
\href{https://doi.org/10.1103/PhysRevLett.116.061102}{doi:10.1103/PhysRevLett.116.061102};
\href{https://arxiv.org/abs/1602.03837}{arXiv:1602.03837}.

\bibitem{Pretorius2005}
F.~Pretorius,
``Evolution of Binary Black-Hole Spacetimes,''
Phys.\ Rev.\ Lett.\ \textbf{95}, 121101 (2005).
\href{https://doi.org/10.1103/PhysRevLett.95.121101}{doi:10.1103/PhysRevLett.95.121101};
\href{https://arxiv.org/abs/gr-qc/0507014}{arXiv:gr-qc/0507014}.



\bibitem{Blanchet2014}
L.~Blanchet,
``Gravitational Radiation from Post-Newtonian Sources and Inspiralling Compact Binaries,''
Living Rev.\ Relativ.\ \textbf{17}, 2 (2014).
\href{https://doi.org/10.12942/lrr-2014-2}{doi:10.12942/lrr-2014-2};
\href{https://arxiv.org/abs/1310.1528}{arXiv:1310.1528}.

\bibitem{Bern2019}
Z.~Bern, C.~Cheung, R.~Roiban, C.-H.~Shen, M.~P.~Solon, and M.~Zeng,
``Scattering Amplitudes and the Conservative Hamiltonian for Binary Systems at Third Post-Minkowskian Order,''
Phys.\ Rev.\ Lett.\ \textbf{122}, 201603 (2019).
\href{https://doi.org/10.1103/PhysRevLett.122.201603}{doi:10.1103/PhysRevLett.122.201603};
\href{https://arxiv.org/abs/1901.04424}{arXiv:1901.04424}.

\bibitem{BarackPound2018}
L.~Barack and A.~Pound,
``Self-force and radiation reaction in general relativity,''
Rep.\ Prog.\ Phys.\ \textbf{82}, 016904 (2019).
\href{https://iopscience.iop.org/article/10.1088/1361-6633/aae552}{doi:10.1088/1361-6633/aae552};
\href{https://arxiv.org/abs/1805.10385}{arXiv:1805.10385}.

\bibitem{Pratten2020}
G.~Pratten, \emph{et al.},
``Computationally efficient models for the dominant and subdominant harmonic modes of precessing binary black holes,''
Phys.\ Rev.\ D \textbf{103}, 104056 (2021).
\href{https://doi.org/10.1103/PhysRevD.103.104056}{doi:10.1103/PhysRevD.103.104056};
\href{https://arxiv.org/abs/2004.06503}{arXiv:2004.06503}.

\bibitem{Field2014}
S.~E.~Field, C.~R.~Galley, J.~S.~Hesthaven, J.~Kaye, and M.~Tiglio,
``Fast Prediction and Evaluation of Gravitational Waveforms Using Surrogate Models,''
Phys.\ Rev.\ X \textbf{4}, 031006 (2014).
\href{https://doi.org/10.1103/PhysRevX.4.031006}{doi:10.1103/PhysRevX.4.031006};
\href{https://arxiv.org/abs/1308.3565}{arXiv:1308.3565}.

\bibitem{Blackman2017}
J.~Blackman \emph{et al.},
``Numerical relativity waveform surrogate model for generically precessing binary black hole mergers,''
Phys.\ Rev.\ D \textbf{96}, 024058 (2017).
\href{https://doi.org/10.1103/PhysRevD.96.024058}{doi:10.1103/PhysRevD.96.024058};
\href{https://arxiv.org/abs/1705.07089}{arXiv:1705.07089}.

\bibitem{Raissi2019}
M.~Raissi, P.~Perdikaris, and G.~E.~Karniadakis,
``Physics-informed neural networks: A deep learning framework for solving forward and inverse problems involving nonlinear partial differential equations,''
J.\ Comput.\ Phys.\ \textbf{378}, 686 (2019).
\href{https://doi.org/10.1016/j.jcp.2018.10.045}{doi:10.1016/j.jcp.2018.10.045};
\href{https://arxiv.org/abs/1711.10561}{arXiv:1711.10561}.

\bibitem{Greydanus2019}
S.~Greydanus, M.~Dzamba, and J.~Yosinski,
``Hamiltonian neural networks,''
arXiv:1906.01563 (2019).
\href{https://arxiv.org/abs/1906.01563}{arXiv:1906.01563}.

\bibitem{Karniadakis2021}
G.~E.~Karniadakis, I.~G.~Kevrekidis, L.~Lu, P.~Perdikaris, S.~Wang, and L.~Yang,
``Physics-informed machine learning,''
Nat.\ Rev.\ Phys.\ \textbf{3}, 422 (2021).
\href{https://doi.org/10.1038/s42254-021-00314-5}{doi:10.1038/s42254-021-00314-5}.

\bibitem{Zhou:2026haz}
Y.~C.~Zhou, H.~Ma, Z.~Cao, T.~Wu, H.~B.~Jin, X.~Feng, S.~Huang, Z.~C.~Zhao and Y.~L.~Wu,
``Solving Hamiltonian constraint equation with physics-informed neural networks,''
Phys. Rev. D \textbf{114}, 023032 (2026).
\href{https://doi.org/10.1103/619s-p6md}{doi:10.1103/619s-p6md};
\href{https://arxiv.org/abs/2607.06002}{arXiv:2607.06002}.

\bibitem{Bayona2026}
E.~Bayona and H.~Quevedo,
``Solving Einstein's Vacuum Equations with Physics-Informed Neural Networks: Boundary Conditions and Domain Decomposition,''
arXiv:2608.08846 (2026).
\href{https://arxiv.org/abs/2608.08846}{arXiv:2608.08846}.

\bibitem{Cranganore:2025vsu}
S.~S.~Cranganore, A.~Bodnar, A.~Berzins and J.~Brandstetter,
``Einstein Fields: A Neural Perspective To Computational General Relativity,''
arXiv:2507.11589 (2025).
\href{https://arxiv.org/abs/2507.11589}{arXiv:2507.11589}.

\bibitem{Yoo:2025yhl}
J.~Yoo, M.~Boyle and N.~Deppe,
``Learning post-Newtonian corrections from numerical relativity,''
Phys. Rev. D \textbf{113}, 084008 (2026).
\href{https://doi.org/10.1103/8tnr-mtb7}{doi:10.1103/8tnr-mtb7};
\href{https://arxiv.org/abs/2511.10522}{arXiv:2511.10522}.

\bibitem{GomesDaSilva2026}
L.~J.~Gomes Da Silva,
``Introducing SINFONIA: Symplectic, slimplectic and Magnusian (Neural) Flows for Orbital Numerical Integration and Acceleration,''
arXiv:2609.03329 (2026).
\href{https://arxiv.org/abs/2609.03329}{arXiv:2609.03329}.

\bibitem{PRLcompanion}
G.~Barbagallo and J.~Matulich,
``Learning Post-Newtonian Dynamics with Gravity-Informed Neural Networks,''
submitted to Phys.\ Rev.\ Lett.\ (2026).

\bibitem{GravINNscode}
J.~Matulich and G.~Barbagallo,
\texttt{GravINNs}: Physics-informed neural networks for post-Newtonian compact-binary dynamics, version v1.0.0, Zenodo (2026),
\href{https://doi.org/10.5281/zenodo.23225699}{doi:10.5281/zenodo.23225699}.

\bibitem{Buonanno:1998gg}
A.~Buonanno and T.~Damour,
``Effective one-body approach to general relativistic two-body dynamics,''
Phys. Rev. D \textbf{59}, 084006 (1999).
\href{https://doi.org/10.1103/PhysRevD.59.084006}{doi:10.1103/PhysRevD.59.084006};
\href{https://arxiv.org/abs/gr-qc/9811091}{arXiv:gr-qc/9811091}.

\bibitem{DJS2000}
T.~Damour, P.~Jaranowski, and G.~Sch\"afer,
``Dynamical invariants for general relativistic two-body systems at the third post-Newtonian approximation,''
Phys.\ Rev.\ D \textbf{62}, 044024 (2000).
\href{https://doi.org/10.1103/PhysRevD.62.044024}{doi:10.1103/PhysRevD.62.044024}.

\bibitem{DJS2001}
T.~Damour, P.~Jaranowski, and G.~Sch\"afer,
``Dimensional regularization of the gravitational interaction of point masses,''
Phys.\ Lett.\ B \textbf{513}, 147 (2001).
\href{https://doi.org/10.1016/S0370-2693(01)00642-6}{doi:10.1016/S0370-2693(01)00642-6}.

\bibitem{BarbagalloandMatulich}
G.~Barbagallo and J.~Matulich, work in progress.

\bibitem{Tancik2020}
M.~Tancik \emph{et al.},
``Fourier features let networks learn high frequency functions in low dimensional domains,''
Advances in Neural Information Processing Systems \textbf{33} (2020).
\href{https://arxiv.org/abs/2006.10739}{arXiv:2006.10739}.


\bibitem{Lu2021}
L.~Lu, P.~Jin, G.~Pang, Z.~Zhang, and G.~E.~Karniadakis,
``Learning nonlinear operators via DeepONet based on the universal approximation theorem of operators,''
Nat.\ Mach.\ Intell.\ \textbf{3}, 218 (2021).
\href{https://doi.org/10.1038/s42256-021-00302-5}{doi:10.1038/s42256-021-00302-5};
\href{https://arxiv.org/abs/1910.03193}{arXiv:1910.03193}.

\bibitem{Li2021}
Z.~Li \emph{et al.},
``Fourier neural operator for parametric partial differential equations,''
ICLR (2021).
\href{https://arxiv.org/abs/2010.08895}{arXiv:2010.08895}.



\bibitem{Gundlach2005}
C.~Gundlach, G.~Calabrese, I.~Hinder, and J.~M.~Mart\'{\i}n-Garc\'{\i}a,
``Constraint damping in the Z4 formulation and harmonic gauge,''
Class. Quantum Grav. \textbf{22}, 3767 (2005).
\href{https://doi.org/10.1088/0264-9381/22/17/025}{doi:10.1088/0264-9381/22/17/025};
\href{https://arxiv.org/abs/gr-qc/0504114}{arXiv:gr-qc/0504114}.

\bibitem{Lincoln:1990ji}
C.~W.~Lincoln and C.~M.~Will,
``Coalescing Binary Systems of Compact Objects to (Post)5/2 Newtonian Order: Late Time Evolution and Gravitational Radiation Emission,''
Phys. Rev. D \textbf{42}, 1123 (1990).
\href{https://doi.org/10.1103/PhysRevD.42.1123}{doi:10.1103/PhysRevD.42.1123}.

\bibitem{Iyer:1995rn}
B.~R.~Iyer and C.~M.~Will,
``Post-Newtonian gravitational radiation reaction for two-body systems: Nonspinning bodies,''
Phys. Rev. D \textbf{52}, 6882 (1995).
\href{https://doi.org/10.1103/PhysRevD.52.6882}{doi:10.1103/PhysRevD.52.6882}.

\bibitem{WangPerdikaris2022}
S.~Wang, S.~Sankaran, and P.~Perdikaris,
``Respecting causality is all you need for training physics-informed neural networks,''
arXiv:2203.07404 (2022).
\href{https://arxiv.org/abs/2203.07404}{arXiv:2203.07404}.






\end{thebibliography}
\end{document}